\documentclass[
tightenlines,
preprintnumbers,
nofootinbib,
twocolumn,
 amsmath,amssymb]{revtex4-2}
\usepackage{amsmath}
\DeclareMathOperator*{\argmax}{arg\,max}
\DeclareMathOperator*{\argmin}{arg\,min}
\usepackage{amssymb}
\usepackage{bbold}
\usepackage{bm}
\usepackage{color,graphicx}
\usepackage{slashed}
\usepackage{comment}
\usepackage{hyperref}
\usepackage{subcaption}
\usepackage{makecell}
\usepackage[dvipsnames]{xcolor}
\usepackage{pifont}
\usepackage{caption}
\usepackage{array}
    \newcolumntype{P}[1]{>{\centering\arraybackslash}p{#1}}
    \newcolumntype{M}[1]{>{\centering\arraybackslash}m{#1}}

\begin{document}

\preprint{ANL-186490}

\title{Learning PDFs through Interpretable Latent Representations in Mellin Space}

\author{Brandon Kriesten}

\author{T.~J.~Hobbs}

\affiliation{
    High Energy Physics Division, Argonne National Laboratory, Lemont, IL 60439
}

\date{\today}

\begin{abstract}
Representing the parton distribution functions (PDFs) of the proton and other hadrons through
flexible, high-fidelity parametrizations has been a long-standing goal of particle physics phenomenology.
This is particularly true since the chosen parametrization methodology can play an influential
role in the ultimate PDF uncertainties as extracted in QCD global analyses; these, in turn, are often determinative of the reach of
experiments at the LHC and other facilities to non-standard physics, including at large $x$, where
parametrization effects can be significant.
In this study, we explore a series of encoder-decoder machine-learning (ML) models with various neural-network
topologies as efficient means of reconstructing PDFs from meaningful information stored in
an interpretable latent space.
Given recent effort to pioneer synergies between QCD analyses and lattice-gauge calculations,
we formulate a latent representation based on the behavior of PDFs in Mellin space, {\it i.e.}, their
integrated moments, and test the ability of various models to decode PDFs from this information
faithfully.
We introduce a numerical package, \texttt{PDFdecoder}, which implements several encoder-decoder models to reconstruct PDFs with high fidelity and use this tool to explore strengths and pitfalls of neural-network approaches
to PDF parametrization.
We additionally dissect patterns of learned correlations between encoded Mellin moments and reconstructed
PDFs which suggest opportunities for further improvements to ML-based approaches to PDF parametrizations
and uncertainty quantification.
\end{abstract}

\maketitle

\section{Introduction}
\label{sec:intro}
An overarching theme of modern high-energy physics (HEP) has been the quest to identify
the microscopic nature of dark matter suggested by astrophysical and related
observations. While this effort has been characterized by an array of (in)direct
detection experiments, it is further complemented by a large battery of standard model (SM)
tests at various colliders, especially the (HL-)LHC~\cite{Apollinari:2017lan}.
The reach of these (hadron) collider-based searches for physics beyond the Standard Model (BSM)
is generally limited by current knowledge of the proton's collinear
structure; for measurements at the LHC, the most relevant quantities are the
unpolarized parton distribution functions (PDFs) of the proton~\cite{Amoroso:2022eow}, which determine the
parton-level luminosities informing cross-section predictions for Higgs,
gauge-boson, or other production processes.

While high-energy measurements are frequently dominated by the low-$x$ behavior
of the PDFs, the rapidity and invariant-mass distributions that serve
as fertile ground for BSM searches can also be significantly dependent on the
very high-$x$ PDFs as well~\cite{ZEUS:2019cou,Carrazza:2019sec,Greljo:2021kvv,Madigan:2021uho,Gao:2022srd,CMS:2021yzl,Iranipour:2022iak,Fu:2023rrs,Ablat:2023tiy}.
For this reason, considerable attention has been trained on the determination
of PDFs at high $x\! \gtrsim\! 0.1$, where sensitive data are sparse~\cite{Jing:2023isu,Courtoy:2020fex,Hobbs:2019gob,Wang:2018heo}
and the parametrization dependence on the chosen functional form used to fit
the PDFs can represent a significant effect which must be quantified and controlled~\cite{Kotz:2023pbu}.
Owing to such considerations, a number of analysis efforts have turned to ML-based
methods like feed-forward neural networks.

In particular, neural networks have been adopted for parametrizing the $x$ dependence
of PDFs in global fits \cite{NNPDF:2021njg}; this is generally on the contention that such approaches
afford enhanced model independence and are unbiased since they do not implement a
specific functional form for the PDFs' $x$ dependence.
While neural networks do indeed represent a highly flexible approach, they also possess
methodological dependences analogous to those in other fitting frameworks, and can be
statistically challenging to sample comprehensively~\cite{Courtoy:2022ocu}; these issues are reflected in
various ways, including the role of network architecture and
procedural choices made during training.
In this study, we investigate such questions using a simplified toy model for the PDFs,
and focus on PDF reconstruction as a self-contained problem which allows a quantitative
assessment of the performance of specific NN architectures and hyperparameter choices.
A natural choice for this problem is that of the {\it autoencoder}, in that this broad class of
machine-learning models entails going from
\begin{equation}
[\mathrm{PDF}] \to \{\mathrm{latent}\} \to [\mathrm{PDF}'] \nonumber
\end{equation}
and thus consist of a $[\mathrm{PDF}]\! \to\! \{\mathrm{latent}\}$ {\it encoding} structure,
followed by a corresponding {\it decoder}, $\{\mathrm{latent}\}\! \to\! [\mathrm{PDF}']$,
as depicted schematically in the first row of
Tab.~\ref{tab:architectures}, which represents the most generic autoencoder (AE)
architecture. Beyond these base ingredients, various potential elaborations
of the associated network structure are depicted in Tab.~\ref{tab:architectures}.

Autoencoder-based models have been used in many contexts in particle physics, including
anomaly detection in experimental measurements~\cite{Farina:2018fyg,Cheng:2020dal,Finke:2021sdf,Ngairangbam:2021yma};
experimental searches for possible BSM signatures~\cite{Cerri:2018anq,Govorkova:2021utb};
fundamental theory~\cite{Hashimoto:2019bih,Hao:2022zns}; and
inverse problems in QCD global analyses~\cite{Almaeen:2022ifg,9534012}.
In the present study, we perform a comprehensive assessment of the capabilities of
AE-based and adjacent ML methods in flexibly representing the PDFs, concentrating
on the flavor-$\mathrm{SU}(2)$ PDF combinations as a smaller-scale toy problem, without significant
loss of generality.
Specifically, we perform a series of PDF reconstructions within a
suite of encoder-decoder network topologies we collectively bundle into a
new computational framework, \texttt{PDFdecoder}. We intend the toy
calculation as an initial demonstration of a general approach that might be
extended to a range of similar problems, including more direct interfaces
with fitted (phenomenological) PDFs.

We connect this first exercise to another active area of PDF phenomenology:
the effort to compute information sensitive to the PDFs' high-$x$ behavior through
lattice gauge techniques~\cite{Hobbs:2019gob,Bringewatt:2020ixn,Hou:2022onq}. As discussed further below, this activity has typically
concentrated either on the evaluation of PDF Mellin moments~\cite{Detmold:2001dv}, as well as direct
information on the $x$-dependent PDFs via the quasi-~\cite{Ji:2014gla} and pseudo-PDF~\cite{Radyushkin:2017cyf} methods (for a comprehensive study of lattice QCD methods for PDF determination see Refs.~\cite{Lin:2017snn,Constantinou:2020hdm}). In the
current work, we concentrate on the former (Mellin moment) approach.
The specific questions we pursue are: ({\it i}) how AEs perform in the generic task of
flexibly parametrizing PDFs; and ({\it ii}) where the limits in the interpretability
of this ML task lie, especially once augmented with hypothetical information on the
underlying PDFs' Mellin moments.
The question of interpretability in this context is multifaceted, and we explore the
performance of various AE architectures in the general PDF reconstruction problem
while simultaneously peering inside the latent representation of various architectures.
We explore the imposition of physics-motivated constraints to the latent space by
requiring the latent space to have an explicit interpretation in terms of the PDFs'
lattice-calculable Mellin moments, and investigate how this requirement influences
patterns of correlations across the various AE network topologies.

Our calculation therefore represents a PDF-specific instance of ongoing work to
develop ML models which physics assumptions directly incorporated into the structure
of trained networks~\cite{Smidt:2020tuy,Desai:2021wbb}.
Moreover, this analysis builds upon recent work to learn the association between PDFs and their
corresponding log-likelihood functions~\cite{Liu:2022plj}, an approach which can be
used in various tasks, including the construction of rapid joint fits of PDFs and
BSM-sensitive SMEFT parameters~\cite{Gao:2022srd} as done in the CTEQ-TEA framework.

The remainder of our paper is as follows: in Sec.~\ref{sec:background}, we discuss
the formal aspects that motivate this work, including key features of QCD (Sec.~\ref{sec:back_theory}
as well as those related to the deployment and use
of autoencoder-based ML models (Sec.~\ref{sec:back_ae}). Sec.~\ref{sec:models} provides
an in-depth discussion of technical features of the ML models used in this work; these
include the particular network configurations we analyze in the case of
autoencoders with discrete latent spaces (in Sec.~\ref{sec:networks}) and those
with variation architectures, as discussed in Sec.~\ref{sec:variational}. In addition,
we summarize general aspects of AEs and variational models from the perspective
of uncertainty quantification (UQ) in Sec.~\ref{sec:UQ} and dimensionality reduction
in Sec.~\ref{sec:dimred}.

In Sec.~\ref{sec:reco}, we outline our main results --- these include the specific use of
the AEs to reconstruct the PDFs from input distributions (\ref{sec:data_param}); the fidelity of the resulting
reconstructions (\ref{sec:PDF-recon}); and the capacity of AEs and VAIMs to serve as generative models to
produce phenomenological PDFs from input moments (\ref{sec:pheno-PDF}); and a quantitative discussion of the
correlations induced by this approach between the inputs moments and reconstructed PDFs (\ref{sec:corr}).
In Sec.~\ref{sec:conclusions}, we highlight several key conclusions.
Lastly, in App.~\ref{sec:pca}, we provide a formal discussion of the relationship between
AEs as explored in our main study and PCA as an alternative dimensionality reduction algorithm.

\section{Background and Formalism}
\label{sec:background}
In this section, we review the theoretical foundations of our study, focusing on both QCD-related aspects as well
as the fundamentals of the ML frameworks used in our calculations. In particular, we first review the relation between
PDFs and their Mellin transforms, which provides a basis to reconstruct the former given knowledge of the PDFs' integrated
moments. We specially highlight this connection as it forms the basis for imposing some interpretable structure on the ML
models explored later in this study. Following this discussion, we summarize the basic formalism for constructing and
training the autoencoder and variational architectures explored in this study. We further elaborate upon the specialization
of each of these network structures to the PDF representation problem in subsequent sections.

\subsection{QCD Theory}
\label{sec:back_theory}

The unpolarized collinear PDFs are the quantum correlation functions which encode the longitudinal structure of the nucleon. These non-local, forward matrix elements parameterize the inherently nonperturbative distribution of quarks and gluons inside the proton; on the basis of QCD factorization~\cite{Collins:2011zzd}, they can be related to observables in physical scattering processes such as the deep-inelastic scattering (DIS) structure functions through convolutions with short-distance Wilson coefficient functions, $C_i$, schematically:
\begin{equation}
    F_i(x,Q) =C_{i, f} \left( {x \over \xi}, {Q \over \mu} \right) \otimes q_f(\xi, \mu)\ .
\label{eq:fact}
\end{equation}
In the expression above, the flavor-dependent PDFs, $q_f$ have been explicitly separated 
from the perturbatively calculable matrix elements at a factorization scale, $\mu$; at 
leading-order in $\alpha_s$, $C\! \sim\! \delta (1\! -\! x/\xi)$, and we consider 
parametrizations of the PDFs, $q_f(x)$, in the case of specific $f\! \in\! \mathrm{SU}(2)$ 
flavor combinations for the remainder of this study.
From Eq.~(\ref{eq:fact}), the enterprise of extracting PDFs from data can be seen as
involving a large-scale inverse problem. In particular,
QCD global analyses typically parametrize the PDFs flexibly at an initial boundary
scale of the perturbative evolution, $\mu\! =\! Q_0$, via
\begin{equation}
    q_f(x,\mu=Q_0) = N_f\, x^{\alpha_f} (1-x)^{\beta_f} \cdot \mathcal{P}(x)\ ,
\end{equation}
in which $\mathcal{P}(x)$ might be a high-order polynomial (dependent on some
array of shape parameters) or a feed-forward neural network, $\mathcal{P}(x)\! =\! \mathrm{NN}(x)$.
The parameters of the polynomial or neural network can be determined
from available hadronic data up to knowledge of the underlying perturbative QCD and
electroweak theory ({\it e.g.}, for the matrix elements, $C_{i, f}$) as well as consideration of many
other theoretical issues with can influence predictions for empirical data.

Adequately constraining the complex shape of the PDFs demands a wide range of data spanning many
processes and energy scales, especially given the need to unravel the flavor dependence
and local $x$ dependence.
Given the theoretical and computational complexity of this task, a series of ongoing
PDF benchmarking efforts have played out~\cite{Jing:2023isu,PDF4LHCWorkingGroup:2022cjn,Butterworth:2015oua,Ball:2012wy},
with special focus on the challenge of obtaining reproducible and understandable PDF uncertainties across various
fitting efforts.

Complementary to these activities, it has also long been understood that the PDFs can be related
to their underlying $x$-integrated moments via an inverse-Mellin transform; in particular,
\begin{equation}
\label{eq:Mellin}
q(x)+(-1)^{n+1}\,\overline{q}(x)=\frac{1}{2\pi i}\int_{c-i\infty}^{c+i\infty}dn\,x^{-n-1}\, \langle x{}^{n}\rangle_{q}\ .
\end{equation}
As such, full knowledge of the moments could allow a complete determination of the corresponding $x$-dependent shapes of the
PDFs, including at very high $x$.
Of course, from Eq.~(\ref{eq:Mellin}) it can be seen that full knowledge requires an infinite tower of Mellin
moments to specify the PDFs over all $x$ and for all flavors. In lieu of such comprehensive information, however,
even partial knowledge of select Mellin moments may supplement experimental constraints.

This realization has motivated an array of nonperturbative methods, especially involving lattice QCD, to calculate the
PDFs' Mellin moments, which can be evaluated as the operators of local matrix elements.
For example, by the operator product expansion (OPE), moments of twist-2 quark PDFs may be related to quark-level operator
insertions evaluated in an appropriate hadronic basis as
\begin{align}
\label{eq:OPE_unp}
2\langle x^n\rangle_{q}\,[p_{\mu_{1}}\cdots &p_{\mu_{n+1}}-{\rm traces}] \nonumber \\
&=\frac{1}{2}\sum_{s}\langle p,s|\mathcal{O}_{\{\mu_{1},\cdots,\mu_{n+1}\}}^{q}|p,s\rangle\ ,
\end{align}
where higher moments correspond to the matrix elements of additional covariant derivatives,
\begin{equation}
	\mathcal{O}_{\{\mu_{1},\cdots,\mu_{n+1}\}}^{q} = \ensuremath{{\displaystyle \bar{q}(x)\, \gamma_{\{\mu_{1}}i\overleftrightarrow{D}_{\mu_{2}}...i\overleftrightarrow{D}_{\mu_{n+1}\}}\, q(x)}}\ .
\label{eq:ops}
\end{equation}
Matrix elements of these latter operators are calculable through lattice QCD techniques for the lowest few
moments, {\it e.g.}, $n\! \le\! 3$. Beyond this order, computational artifacts cloud the signal-to-noise
needed for a reliable extraction of the moments.

Through recent lattice developments~\cite{Bhattacharya:2023ays,Bhattacharya:2023jsc,Bhattacharya:2022aob}, it is also
possible to compute generalized parton distributions (GPDs) \cite{Ji:1996ek, Muller:1994ses,Radyushkin:1997ki} and their
associated moments at finite momentum transfer, $t\! =\! (p\!-\!p')^2$, relevant for exclusive processes like deeply virtual
Compton scattering (DVCS) \cite{Ji:1996nm}; the resulting moments at finite $t$ may then be extrapolated to $t\!=\!0$ assuming
a parametric form for the $t$ dependence.
This procedure thus has the potential to provide further access to higher Mellin moments
in a way that complements the direct lattice calculations corresponding to Eq.~(\ref{eq:OPE_unp}).
In particular, the lowest few ($x$-integrated) Mellin moments of the unpolarized quark-GPDs can
be expressed as linear combinations of $t$-dependent matrix elements weighted by the skewness, $\xi$ \cite{Ji:1997gm}:
\begin{eqnarray}
    \int_{-1}^{1}dx x H^{q}(x,\xi,t) &=& A^{q}_{2,0}(t) + (2\xi)^{2} C^{q}_{2}(t)\ , \nonumber \\
    \int_{-1}^{1}dx x^{2}H^{q}(x,\xi,t) &=& A^{q}_{3,0}(t) + (2\xi)^{2}A^{q}_{3,2}(t)\ , \nonumber \\
    \int_{-1}^{1}dx x^{3} H^{q}(x,\xi,t) &=& A^{q}_{4,0}(t) + (2\xi)^{2}A^{q}_{4,2}(t) \nonumber \\
                                         &+& (2\xi)^{4}C^{q}_{4}(t)\ , \nonumber \\
    \int_{-1}^{1}dx x^{4}H^{q}(x,\xi,t) &=& A^{q}_{5,0}(t) + (2\xi)^{2}A^{q}_{5,2}(t) \nonumber \\
                                        &+& (2\xi)^{4}A^{q}_{5,4}(t)\ .
\end{eqnarray}
In the forward ($\xi\! \to\! 0$) limit, these expressions simplify to direct relations for the conventional
PDF Mellin moments in terms of lattice-calculable GPD matrix elements; namely,
\begin{eqnarray}
    \langle x^{n} \rangle_{q} &=&  \int_{-1}^{1}dx x^{n} H^{q}(x,0,0) \nonumber \\
                              &=& \int_{-1}^{1}dx x^{n} q(x)  = A_{n+1,0}(0)\ , 
\end{eqnarray}
which may be rendered over the proper region of support, $x\! \in\! [0,1]$:
\begin{eqnarray}
\label{eq:moment}
    \langle x^{n} \rangle_{q} &=& \int_{0}^{1} dx\, x^n \Big[ q(x) + (-1)^{n+1}\bar{q}(x) \Big]\ .
\end{eqnarray}
As reflected in the expression above, we note that successive Mellin moments are sensitive to PDF combinations
with an alternating even and odd structure under $C$ symmetry; that is, the lattice-accessible moments are, for
the natural numbers, $n\! =\! 0, 1, \cdots$,
\begin{align}
    &\langle x^{2n}\rangle_{q^-} = \int_{0}^{1} dx x^{2n}\, \left[ q(x) - \bar{q}(x) \right] \nonumber \\
    &\langle x^{2n+1}\rangle_{q^+} = \int_{0}^{1} dx x^{2n+1}\, \left[ q(x) + \bar{q}(x) \right]\ ,
\label{eq:even-odd-moments}
\end{align}
due to the action of covariant derivatives on the quark fields of Eq.~(\ref{eq:ops}). As an example, for the $u$-PDFs
explored in this study, the relevant moments are thus $\langle 1 \rangle_{u^-}$, $\langle x \rangle_{u^+}$,
$\langle x^2 \rangle_{u^-}$, $\cdots$, and similarly for the $d$-PDFs.
We stress this point as we will exploit this mathematical structure to constrain some of the latent
spaces of the networks introduced in Sec.~\ref{sec:back_ae} below, particularly, those of the AE-CL, AE-WC, and VAIM models.
For a complete review of the limits and properties of PDF Mellin moments, GPDs, and
interrelations among these, see, {\it e.g.}, Refs.~\cite{Lin:2017snn,Constantinou:2020hdm,Diehl:2003ny}.

\subsection{Encoder/Decoder Networks Theory}
\label{sec:back_ae}

\subsubsection{Autoencoders}

An autoencoder (AE)~\cite{37f2b6bee745402aa4e4d124d33be0e0,Bourlard1988AutoassociationBM,Goodfellow-et-al-2016} is a system
of two feed-forward neural networks constructed from multi-layer perceptrons that work to encode input data into a compressed,
informative representation of the feature space and then decode it. The goal of an AE is to construct a reduced, and as a
consequence, lossy, representation of the input space such that it retains as much of the input variance as possible while
maintaining global properties of the data. Together, a fully trained system of encoder and decoder networks approximate a
unitary transformation as applied to the input data. 

The encoder network maps input data, $\vec{x} \in \mathbb{R}^{m}$, to a compressed feature subspace, $\vec{z} \in \mathbb{R}^{n}$,
where $n\! <\! m$; conversely, the decoder network takes as its input $\vec{z} \in \mathbb{R}^{n}$ and up-samples the feature
space back to $\mathbb{R}^{m}$.
Mathematically, the encoder network, $e_{\theta}(x)$, and decoder network, $d_{\phi}(z)$, are layered series of affine
transformations constructed from parameter sets we label $\theta$ and $\phi$, respectively, followed by element-wise,
nonlinear activation functions. The fully-connected layers feed forward with the $(k+1)^\mathit{th}$ layer parametrically
determined from the $k^\mathit{th}$ as
\begin{eqnarray}
    x_{k+1} = A_{k}\big( W_{k}\, x_{k} + b_{k} \big)\ ,
\end{eqnarray}
where $W_k$ and $b_k$ are the weight matrix and bias vector, respectively, and $A_{k}$ is the above-mentioned nonlinear
activation function; this can assume a variety of nonlinear forms but is commonly chosen to be the ReLU or ramp function,
$\max{(0,x)}$, whose derivative is discontinuous and sets negative outputs to 0 \cite{agarap2019deep}; variations on ReLU
are also possible, for instance, the ELU function, which is similar but instead sets negative ($x\! <\! 0)$ inputs to
$\alpha ( e^{x}\! -\! 1)$, where $\alpha$ is a positive semi-definite hyperparameter which might be varied over iterative
training(s)~\cite{clevert2016fast}.

The reconstruction loss can be monitored with a loss metric that is conducive to the particular problem, in this case we
use mean-squared error defined as:
\begin{eqnarray}
    \mathcal{L}_{\theta,\phi} = \frac{1}{N} \Big \| x - d_{\phi}[e_{\theta}(x)] \Big \|^{2}_{2}\ .
\label{eq:recon-loss}
\end{eqnarray}
where $\| \cdot \|_{2}$ denotes the Euclidean norm of the vector and $N$ is the total number of elements in the vector. The
output of the encoder network --- the \textit{latent vector}, $\vec{z}$, or \textit{latent space} --- is a representation
of the input features in a compressed subspace. The autoencoder configures $\vec{z}$ such that the decoded latent vector
is as close to the original as possible:
 \begin{eqnarray}
     x \approx d_{\phi}[e_{\theta}(x)]\ .
 \end{eqnarray}
In performing this task, the latent space becomes encoded into an ultimate configuration which is a formally intractable
distribution; as such {\it there is no way to construct this distribution in a meaningful way outside of the use of the
encoder network}. Since the representation of the latent space is intractable, the input data must be passed from encoder
to decoder in order to properly perform the compression and subsequent reconstruction. 

The encoding of the latent vector is similar to that of the projection of a feature space onto some subspace using principal
component analysis (PCA); in fact, the goals of both unsupervised learning techniques are the same. As discussed further in
Sec.~\ref{sec:dimred} and explicitly demonstrated in App.~\ref{sec:pca}, the methods are identical when the encoder and decoder
networks in an AE are constructed from a single linear transformation. The choice of encoder and decoder is like the calculation
of the principal components from the covariance matrix, with the end goal being the encapsulation of as much variance in the
data set as possible while simultaneously ensuring that global properties such as Euclidean distance are preserved. In PCA, the
largest eigenvalues of the covariance matrix satisfy these conditions, while in an AE,  $e_{\theta}^{*}$ and $d_{\phi}^{*}$ are
chosen from all encoder and decoder networks, $E$ and $D$, so as to perform the minimization task,
\begin{eqnarray}
e_{\theta}^{*},d_{\phi}^{*} = \argmin_{e_{\theta}\in E, d_{\phi} \in D}\Big(x - d_{\phi}[e_{\theta}(x)]\Big)\ .   
\end{eqnarray}
As in PCA, the AE encodes data at a loss, meaning that the decoding process will never fully reconstruct the input data with
100\% accuracy.
 
AEs have typically been used in transfer learning~\cite{10.5555/2832747.2832823}, (non)linear dimensionality
reduction~\cite{doi:10.1126/science.1127647,WANG2016232}, de-noising inputs~\cite{10.1145/1390156.1390294}, and anomaly
detection~\cite{8363930}, with a number of applications in HEP as noted in Sec.~\ref{sec:intro}. Meanwhile, AEs have
proven difficult to use as generative models due to the latent space intractability highlighted above. Extra caution
should be taken as empty regions in the latent space are often out-of-distribution for the decoder network and therefore
could lead to nonsense generation. Since the shape of the latent space distribution is unregulated, it is generally difficult
to interpolate over empty regions where no encodings lie. Without some constraint on the learned latent representation, there
is typically no way to use the decoder network as a generative algorithm.
It is this fundamental limitation of AEs and ML algorithms as applied to PDF reconstruction
that we confront in this analysis.

\subsubsection{Variational Autoencoders}
\label{sec:vae}

Variational Autoencoders (VAEs)~\cite{kingma2022autoencoding, DBLP:journals/corr/abs-1906-02691} are a solution to a
computationally difficult/intractable problem of variational inference over an unknown posterior to produce a likelihood
distribution. The VAE solves the problem of out-of-distribution sampling from the latent space because VAEs learn distributions
over the output parameters of the encoder network rather than just the output parameters themselves as in the AE case. 

Consider a scenario in which we wish to create samples of data drawn from some likelihood function, $x\! \sim\! p(x | z)$, where
$\vec{x} \in \mathbb{R}^{m}$ and $\vec{z} \in \mathbb{R}^{n}$ as before. The latent samples are drawn from a prior distribution,
$z \sim p(z)$, while the posterior distribution of the latent space given the input data is given by Bayes' rule,
\begin{eqnarray}
    p(z|x) &=& \frac{p(x|z)p(z)}{p(x)}\ .
\end{eqnarray}
Notice that the denominator --- the evidence or marginal likelihood, $p(x)\! =\! \int_{z \sim p(z)} dz\, p(x|z)p(z)$ --- is
typically intractable because it is usually impossible to calculate all combinations of latent samples that make up $p(z)$.
The intractability of $p(x)$ makes it very computationally challenging to fully map the posterior distribution $p(z|x)$.

We can use variational inference to overcome this challenge by constructing an approximate distribution, $q_{\lambda}(z|x)$,
where $\lambda_{i} \in \{\mu_{i},\sigma^{2}_{i}\}$. To ensure that the approximated distribution, $q_{\lambda}(z|x)$, is as
close as possible to the true posterior, $p(z|x)$, we can monitor the KL-divergence between the two distributions:
\begin{widetext}
\begin{eqnarray}
D_{KL}\Big(q_{\lambda}(z|x) \Big\| p(z|x)\Big) &=& \mathbb{E}_{z\sim q_{\lambda}}\Big(\ln{(q_{\lambda}(z|x))}\Big) - \mathbb{E}_{z\sim q_{\lambda}}\Big(\ln{(p(z|x))}\Big) \nonumber \\
&=& \mathbb{E}_{z\sim q_{\lambda}}\Big(\ln{(q_{\lambda}(z|x))}\Big) - \mathbb{E}_{z\sim q_{\lambda}}\Big(\ln{(p(x|z)p(z))}\Big) + \ln(p(x))\ ,
\end{eqnarray}
\end{widetext}
in which the quantity appearing in the second line above,
\begin{align}
\mathcal{L} = -&\mathbb{E}_{z\sim q_{\lambda}}\Big(\ln{(q_{\lambda}(z|x))}\Big) \nonumber\\
+\ &\mathbb{E}_{z\sim q_{\lambda}}\Big(\ln{(p(x|z)p(z))}\Big)\ ,
\end{align}
is the evidence lower bound (ELBO). During training, the ELBO can be maximized, which is equivalent to the minimization of the
desired KL divergence above up to an unknown constant --- the log evidence, $\ln(p(x))$.

Let us consider using two neural networks, one to learn to approximate the posterior distribution --- to encode the input information
into a distribution over latent variables --- and one to approximate the log-likelihood function in order to decode the latent variables
into a distribution over data. We can then rewrite the ELBO using these neural networks:
\begin{align}
    \mathcal{L} &= -D_{KL}\Big( q_{\phi}(z|x)\Big\| p_{\theta}(z)\Big) \nonumber\\
    &+ \mathbb{E}_{z\sim q_{\phi}}\Big(\ln{(p_{\theta}(x|z)} \Big)\ .
\end{align}
The variables $\theta$ encompass the parameters of the generative model, and the variables $\phi$ encompass the variational
parameters. Notice that the ELBO contains two terms which are the learning goals of the VAE: the first term denotes
organizing the latent space as close to the prior distribution as possible, and the second term is the reconstruction
of the input data conditioned on the latent space. The trade-off between maximizing the ELBO and minimizing the KL divergence
is variational inference.

Now that the distribution over the latent variables conditioned on the input data is regularized, we can use the VAE to generate
new samples that the network has not seen. This regularization organizes the latent space such that there are no holes and
minimizes the chance of out-of-distribution sampling.

\begin{table*}[ht] 
  \renewcommand{\arraystretch}{1.25}
  \centering 
  \resizebox{\textwidth}{!}{
  \begin{tabular}{
  |
  P{0.5in}|
  P{2.in}|
  P{1.5in}|
  P{0.7in}|
  P{0.7in}|
  P{0.7in}|
  P{0.7in}|} 
  \hline\hline\vspace{0.5 pt} Name & \vspace{0.5 pt}Diagram & \vspace{0.5 pt}Loss  & Recreates \newline PDFs  & Tractable Latent  & Free \newline Latent \newline Dimension   &  Moment Constraint\\
      \hline\hline\vspace{0.75cm} AE &  \vspace{0.05cm} 
      \includegraphics[width=2in,height=0.75in]{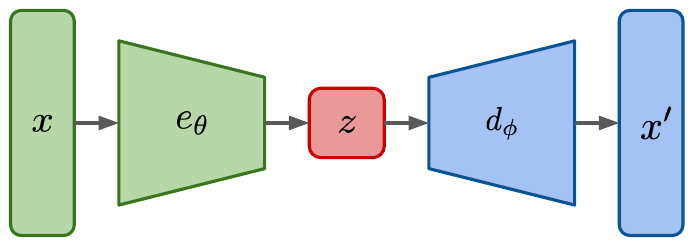} & \vspace{0.05cm} $$\displaystyle \mathcal{L} = \| x - d_{\phi}(e_{\theta}(x)) \|^{2}_2 $$ & \vspace{0.05cm}$$\scalebox{2.5}{\textcolor{ForestGreen}{\checkmark}}$$ & \vspace{0.75cm}\scalebox{2.5}{\textcolor{BrickRed}{\ding{55}}}  & \vspace{0.05cm}$$\scalebox{2.5}{\textcolor{ForestGreen}{\checkmark}}$$ & \vspace{0.75cm}\scalebox{2.5}{\textcolor{BrickRed}{\ding{55}}}\\
      \hline
      \vspace{0.75cm} AE-CL & \vspace{0.05cm} 
      \includegraphics[width=2in,height=0.75in]{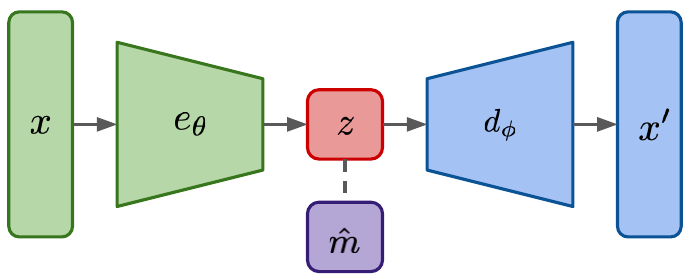} &  \vspace{0.05cm} \begin{eqnarray*}\displaystyle \mathcal{L} &=& \| x - d_{\phi}(e_{\theta}(x)) \|^{2}_2 \nonumber \\
      &+& \| z - \hat{m} \|^{2}_2 \end{eqnarray*} & \vspace{0.05cm}$$\scalebox{2.5}{\textcolor{ForestGreen}{\checkmark}}$$ & \vspace{0.05cm}$$\scalebox{2.5}{\textcolor{Dandelion}{\checkmark}}$$ & \vspace{0.05cm}$$\scalebox{2.5}{\textcolor{Dandelion}{\checkmark}}$$&\vspace{0.05cm}$$\scalebox{2.5}{\textcolor{ForestGreen}{\checkmark}}$$\\
      \hline
       \vspace{0.75cm} AE-WC & \vspace{0.05cm} 
      \includegraphics[width=2in,height=0.75in]{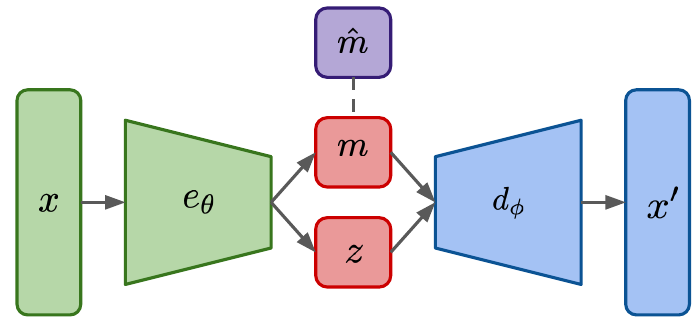} &  \vspace{0.05cm} \begin{eqnarray*}\displaystyle \mathcal{L} &=& \| x - d_{\phi}(e_{\theta}(x)) \|^{2}_2 \nonumber \\
      &+& \| m - \hat{m} \|^{2}_2 \end{eqnarray*} & \vspace{0.05cm}$$\scalebox{2.5}{\textcolor{ForestGreen}{\checkmark}}$$ & \vspace{0.75cm}\scalebox{2.5}{\textcolor{BrickRed}{\ding{55}}} & \vspace{0.05cm}$$\scalebox{2.5}{\textcolor{ForestGreen}{\checkmark}}$$ &\vspace{0.05cm}$$\scalebox{2.5}{\textcolor{ForestGreen}{\checkmark}}$$\\
      \hline
       \vspace{0.75cm} VAE & \vspace{0.05cm} 
      \includegraphics[width=2in,height=0.75in]{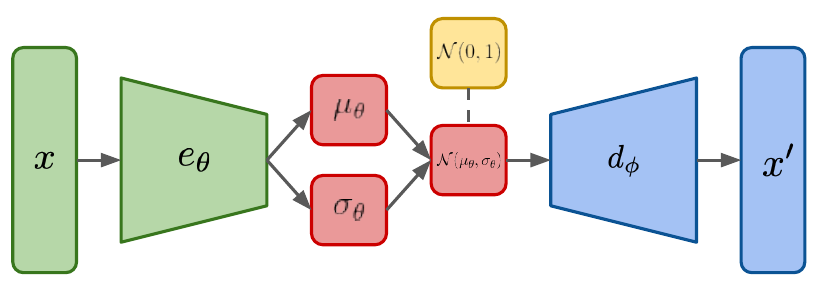} &  \vspace{0.05cm} \begin{eqnarray*}\displaystyle \mathcal{L} &=& \| x - d_{\phi}(e_{\theta}(x)) \|^{2}_2 \nonumber \\
      && \hspace{-1cm} + KL(\mathcal{N}(\mu_{\theta},\sigma_{\theta} | \mathcal{N}(0,1)) \end{eqnarray*} & \vspace{0.05cm}$$\scalebox{2.5}{\textcolor{ForestGreen}{\checkmark}}$$ & \vspace{0.05cm}$$\scalebox{2.5}{\textcolor{ForestGreen}{\checkmark}}$$ & \vspace{0.05cm}$$\scalebox{2.5}{\textcolor{ForestGreen}{\checkmark}}$$ &\vspace{0.75cm}\scalebox{2.5}{\textcolor{BrickRed}{\ding{55}}}\\
      \hline
      \vspace{0.75cm} VAIM & \vspace{0.05cm} 
      \includegraphics[width=2in,height=0.75in]{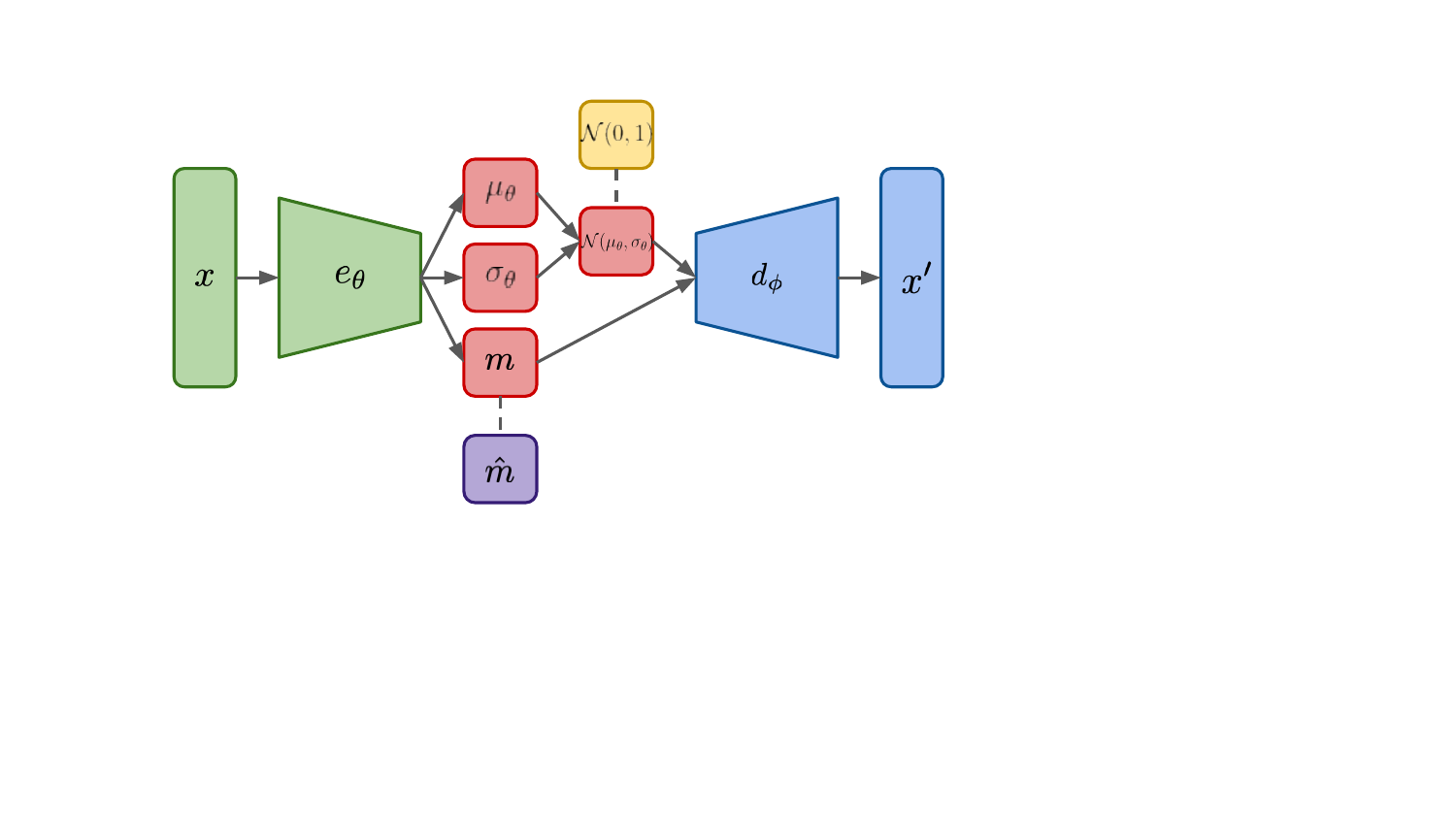} &   \begin{eqnarray*}\displaystyle \mathcal{L} &=& \| x - d_{\phi}(e_{\theta}(x)) \|^{2}_2 \nonumber \\
      &+& \| m - \hat{m} \|^{2}_2 \\
      && \hspace{-1cm} + KL(\mathcal{N}(\mu_{\theta},\sigma_{\theta} | \mathcal{N}(0,1)) \end{eqnarray*} & \vspace{0.05cm}$$\scalebox{2.5}{\textcolor{ForestGreen}{\checkmark}}$$ & \vspace{0.05cm}$$\scalebox{2.5}{\textcolor{ForestGreen}{\checkmark}}$$ & \vspace{0.05cm}$$\scalebox{2.5}{\textcolor{ForestGreen}{\checkmark}}$$ &\vspace{0.05cm}$$\scalebox{2.5}{\textcolor{ForestGreen}{\checkmark}}$$\\
      \hline
      \hline
  \end{tabular}
  }
  \caption{Encoder-decoder network topologies for Mellin-moment latent representations and PDF reconstructions. For each architecture, we
  define the loss surface employed during training and score the corresponding model according to several interpretability criteria for
  (moment) $\to$ (PDF) decodings.} 
  \label{tab:architectures}
\end{table*}

\section{Machine-learning Models}
\label{sec:models}
In this paper, we explore the use of encoder-decoder networks to accomplish the simultaneous goals of PDF reconstruction and de-noising, in concert with transformations from Mellin to $x$ (PDF) space as an additional interpretability task. We consider various topologies of such encoder-decoder frameworks in order to disentangle their differing performance strengths. In Tab.~\ref{tab:architectures}, we summarize each of these architectures in terms of their associated network structure, the loss functions they optimize, and list whether or not they can perform specific tasks. The diagrams graphically parallel the mathematical discussion in Sec.~\ref{sec:back_ae} above, and illustrate how information flows through each network. The feed-forward nature of the networks is represented by arrows, while the dashed lines between network blocks indicate the presence of a loss metric evaluated based on some combined input.

For an architecture to be a robust solution to the problem of mapping Mellin moments to $x$-dependent PDFs, four criteria as summarized in the right columns of Tab.~\ref{tab:architectures} must be satisfied. In particular, the network must ({\it i}) reliably reconstruct the PDF at the final output layer of the architecture; ({\it ii}) it must include a tractable latent for data generation; ({\it iii}) the latent dimensions must be flexible/free; and ({\it iv}) the algorithm must encode the Mellin moments.

On this basis, we examine five unique encoder-decoder network configurations: a generic autoencoder (AE)~\cite{37f2b6bee745402aa4e4d124d33be0e0,Bourlard1988AutoassociationBM}, autoencoders with constrained latent spaces (AE-CL), autoencoders with constraint (AE-WC), variational autoencoders (VAEs)~\cite{kingma2022autoencoding}, and variational autoencoder inverse mappers (VAIM)~\cite{9534012}; all of these entail distinct network topologies broadly within the same encoder-decoder framework. All five network structures are capable of reconstructing PDFs in the final state of the algorithm, corresponding to robust minimization of the reconstruction loss appearing in Eq.~(\ref{eq:recon-loss}). The AE-CL, VAE, and VAIM all have tractable latent spaces, meaning that they can be used for generating new outputs. The AE-CL latent tractability is dependent on the size and comprehensiveness of the training set; this architecture therefore receives a yellow check mark in Tab.~\ref{tab:architectures}, indicating that, in practice, AE-CL models are conditionally tractable in a fashion which much be empirically confirmed. All five architectures possess a free latent dimension to allow increased flexibility for fitting. The AE-CL latent dimension is controlled by the number of moments used; therefore, it is free in theory but in practice is limited by the number of calculated moments. Finally, the AE-CL, AE-WC, and VAIM all use a constraint to the calculated Mellin moments.

We emphasize that, for PDF reconstruction studies, we use {\it undercomplete} autoencoder
architectures --- {\it i.e.}, configurations in which hidden layers are of lower dimensionality
than the initial input layer; we do so in order to constrict PDF information through a bottleneck
in the latent layer in line with our dimensionality reduction and Mellin space interpretability goals. 
The high dimensionality of the initial feature tensor makes this undercomplete architecture advantageous,
as compared to an overcomplete architecture in which the latent layer is dimensionally larger than the initial
feature space.
In all network configurations, it is essential to guarantee balance between the
model complexity of the encoder and the dimensional size of the latent space; doing
so is vital to ensuring interpretability of the encoded latent and to preventing {\it mode collapse},
whereby the entirety of the PDF reconstruction task is absorbed
into network parameters such that a single latent dimension can function as
a label for a diverse range of encoded PDF shapes.
In practice, we actively monitor the network and loss surface during training to
avoid this scenario and optimize the network depth, activation functions, and other
hyperparameters relative to the latent accordingly.

The trade-off between undercomplete {\it vs}.~overcomplete architectures is that, in the undercomplete architecture, forcing information through a bottleneck naturally creates a loss of information; however, regularization of the latent vector is easier since it is of lower dimensionality. On the other hand, overcomplete architectures can memorize and recreate data features (and thus need careful regularization) and can thus reconstruct the PDFs with high fidelity, in principle; however, regularizing a very large-dimensional latent space can complicate training, and we defer such overcomplete models for later study.

In the end, two basic architectures achieve all formal requirements for tractable PDF reconstruction from Mellin space: VAIM, and conditionally, AE-CL models. In Sec.~\ref{sec:reco}, we mainly present PDF generation results obtained with both architectures and further enumerate the advantages and limitations of each approach.
Before doing so, we describe in the remainder of this section general aspects of the PDF implementation in the autoencoder and variational
architectures introduced above as well as issues
in uncertainty quantification and dimensionality reduction.

\subsection{Autoencoder Architectures}
\label{sec:networks}
The AE, AE-CL, and AE-WC have similar architectures constructed of an encoder and decoder network that are trained simultaneously.
The encoder is initiated with an input layer constructed from value arrays taken from four PDF combinations, $[u\pm\bar{u}](x)$ and $[d\pm\bar{d}](x)$, which are logarithmically sampled 196 times from $x_{min} = 10^{-2}$ to $x_{max} = 0.999$. This input layer is followed by five fully connected hidden layers activated by the ELU activation function, where at each fully connected layer we step down the number of nodes by a factor of 2.

The output of the encoder models is a latent vector of varying dimensionality; in this analysis, we assume a baseline dimensionality of 32 (essentially, 8 dimensions per PDF combination) before considering a more highly compressed 8-dimensional latent (2 dimensions per PDF combination). We consider symmetric autoencoder architectures, such that, on the decoder side of the network, the latent vector feeds into five fully connected hidden layers again activated by the ELU activation function, where at each fully connected layer we now step up the number of nodes by a factor of 2 before terminating in
the decoder output, which is a vector containing the reconstructed PDFs.
We use an early stopping algorithm to monitor training and validation loss metrics for overfitting and subsequent memorization of the training data.

\subsection{Variational Architectures}
\label{sec:variational}
\begin{figure}
    \centering
    \vspace{-0.2cm}
    \includegraphics[width=\columnwidth]{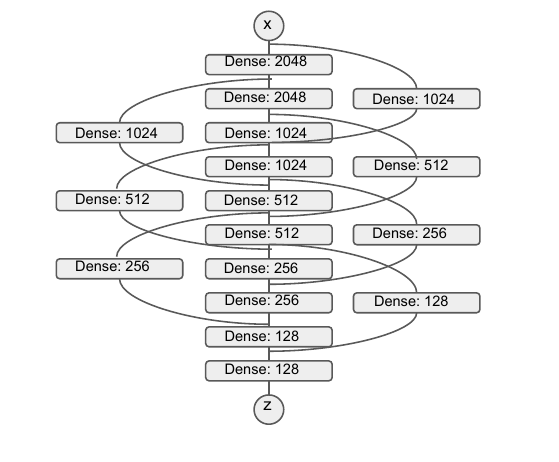} \\
    \vspace{-0.5cm}
    \caption{ResNet-like architecture used for the encoder (pictured) and decoder algorithms of the VAIM and VAE models.}
    \label{fig:resnet}
\end{figure}

The VAE and VAIM models in this analysis both use a ResNet-like architecture~\cite{DBLP:journals/corr/HeZRS15} rather than the simpler feed-forward structure as in the autoencoder architectures. The ResNet algorithm (schematically represented for the encoder model in Fig.~\ref{fig:resnet}) creates shortcut or jump connections between layers in deep neural networks such that gradients survive during backpropagation. Since there are more constraints on the latent and learned information in the VAE and VAIM, these models are modified through the introduction of ResNet layers to ensure robustness and stability. 

Both the VAE and the VAIM models are constructed from fully connected layers in the ResNet-like algorithm with nodes sequentially stepping down from 2048 to 128 by successive factors of 2. The jump connections have a feature transformation such that they can be summed. Each layer is activated through an ELU function, and L2-kernel regularization~\cite{DBLP:journals/corr/abs-1205-2653} is applied at each layer. The corresponding decoder model is a mirror image of the encoder to ensure that both simultaneously perform with comparable fidelity.

\subsection{Uncertainty Quantification}
\label{sec:UQ}
As discussed in Sec.~\ref{sec:intro}, consistent and reproducible PDF uncertainties remain a subject of ongoing effort in HEP. For methods involving neural networks and related ML algorithms, it is essential to quantify uncertainties in model predictions. From a purely computational perspective, many uncertainty quantification (UQ) techniques\footnote{For a comprehensive review on the subject see Ref.~\cite{DBLP:journals/corr/abs-2011-06225} and references therein.} have been developed for the purpose of categorizing sources of uncertainty according to their aleatoric and epistemic nature, using Bayesian statistics to estimate uncertainty, and exploiting sources of uncertainty in neural networks. Some examples of these techniques include Bayesian neural networks (see Ref.~\cite{MacKay1995ProbableNA} and citations therein for original works), Dropout~\cite{srivastava2014dropout} as an approximation to Bayesian inference~\cite{gal2016dropout}, and Dirichlet prior networks~\cite{malinin2018predictive}. The incorporation of these and related uncertainty quantification techniques into HEP and nuclear theory remains at a frontier level. A few examples of investigating UQ techniques in the setting of ML for PDF and GPD analyses are Refs.~\cite{Almaeen:2022imx,Hunt-Smith:2022ugn}.

In the present study, we use the inherent randomness in our neural network architectures to quantify uncertainty through the model ensembling method. First, we ensemble 100 trained models by stochastically varying the random seed that initializes the parameters of the neural network. This ensures that the uncertainty associated with the initialization of the model's parameters is properly folded into the network's predictions. 

On top of this inherent initialization uncertainty, we further exploit the variational aspect in the architecture of these models.
In a variational autoencoder, the latent space is organized with a distribution over the encodings. This distribution is typically taken as a uniform distribution, such that the variance of the normal distribution can be used to make many predictions from the decoder with the same mean.
We leverage this property by including an uncertainty within a given trained model by generating additional 10,000-sample ensembles for each
model configuration, and
incorporate this error as well. We note that this approach to the uncertainty quantification accounts for the potential variability across trained
models rather than computing a bootstrapped uncertainty obtained through consideration of distinct subsets of the full training set.

Finally, in the latent space between the encoder and decoder networks, we predict the Mellin moments of the PDFs. In the AE-CL model, the latent distribution is tied directly to the Mellin moments, whereas they are adjacent to the latent space as a separate output of the encoder. The Mellin moments have some predicted uncertainty which we propagate to obtain a corresponding uncertainty on the predicted output PDFs. These three error sources are combined by randomly sampling within the uncertainties during inference. 

While these uncertainties can be determined for a given model architecture, a broader class of methodological
uncertainties associated with the selected architecture and choice of various hyperparameters is also
relevant and challenging to quantify. This difficulty forms a primary motivation for the present study and 
its examination of an array of model configurations as discussed above; in Sec.~\ref{sec:reco} below,
we highlight the fact that variations in the ML model can be consequential for the resulting predictions
and their interpretability. We reserve a more thorough dissection of these issues for a future study.

\subsection{Dimensionality Reduction}
\label{sec:dimred}
\begin{figure}
    \centering
    \includegraphics[width=\columnwidth]{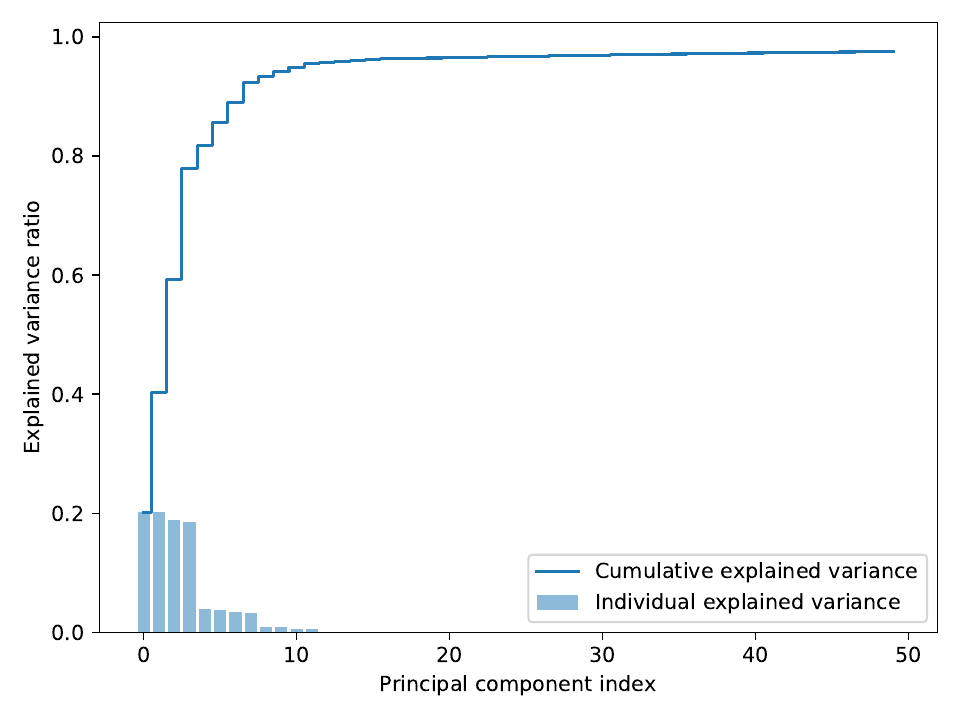}
    \caption{The explained variance ratio from the principal component analysis on the training data of PDFs.}
    \label{fig:pdf_pca}
\end{figure}

Dimensionality reduction techniques can reveal key aspects of a feature data set by mapping
from the original multi-dimensional representation of the data to a more compact space which
captures a maximal degree of the variability of the data.
As such, the encoding algorithms in autoencoders belong to a class of ML-based dimensionality
reduction methods.
At the same time, many dimensionality reduction techniques do not require traditional machine learning, including
principal component analysis (PCA) and singular value decomposition (SVD); we present a brief theoretical
comparison between the autoencoder frameworks described above and PCA in App.~\ref{sec:pca}. A similar mathematical comparison
can be constructed for SVD, although we do not reconstruct this argument here. By the proof given in App.~\ref{sec:pca}
it can be shown that an undercomplete autoencoder with fully connected layers and only linear activation functions has the
exact same loss objective as PCA. To help understand the lower-dimensional latent spaces encoded by our autoencoder methods, in Fig.~\ref{fig:pdf_pca} we plot the explained variance ratio obtained by
performing PCA on our input PDF data set.

Several contiguous groups of 4 principal components with similar values are evident in Fig.~\ref{fig:pdf_pca}, with
each grouping having a diminished value of the explained variance ratio. While the exact nature of the principal
components is unknown owing to the fully unsupervised nature of PCA, by construction these principal components are
some combination of the parameters of the PDF toy model. The three leading groups provide over 90\% of the explained
variance ratio, meaning that any choice of 12 or more principal components would capture most of the information from
the projection; by extension, a 12-dimensional AE latent space is similarly likely to provide strong encoding power
over our original PDF parametrization. We will explore this point explicitly in Sec.~\ref{sec:reco} below.

\section{Results}
\label{sec:reco}
In this section, we present the main results of our analysis of interpretable (Mellin moment) $\to$ (PDF) reconstructions
from the ML models and latent-space organizations described in Sec.~\ref{sec:models}. After discussing the generation of
a training set of PDFs to serve as input at the initial encoder layer, we dissect the performance of the reconstructions
within each model. As a further test, we use our trained model as a generative framework to compare against phenomenological
PDFs; in this context, we explore the statistical correlations between Mellin moments and decoded PDFs in the AE-CL and VAIM
models.

\subsection{Input Layer PDFs}
\label{sec:data_param}
\begin{figure*}[!ht]
    \centering
    \vspace*{-1.3cm}
    \includegraphics[width=0.85\columnwidth]{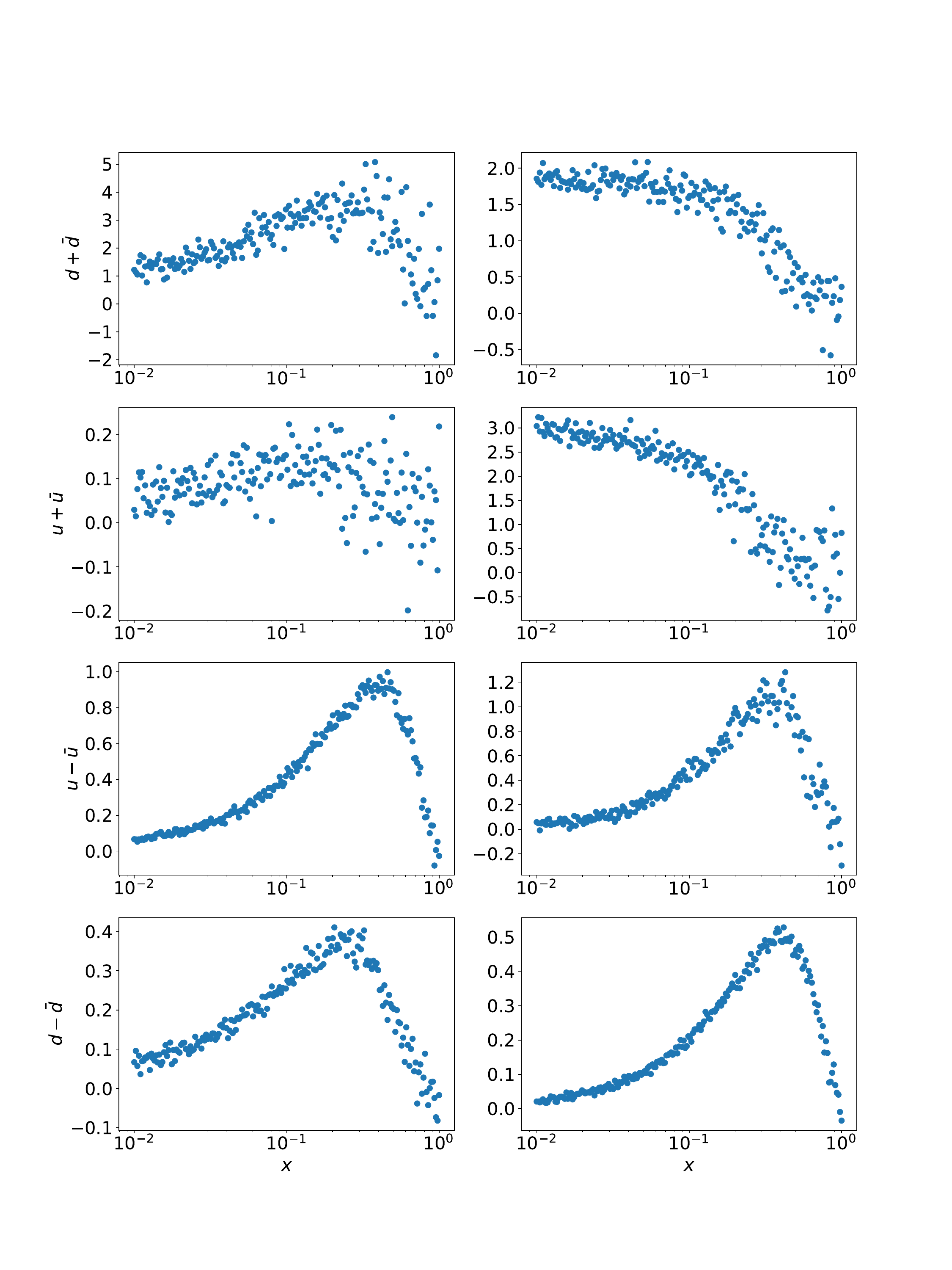}
    \includegraphics[width=0.85\columnwidth]{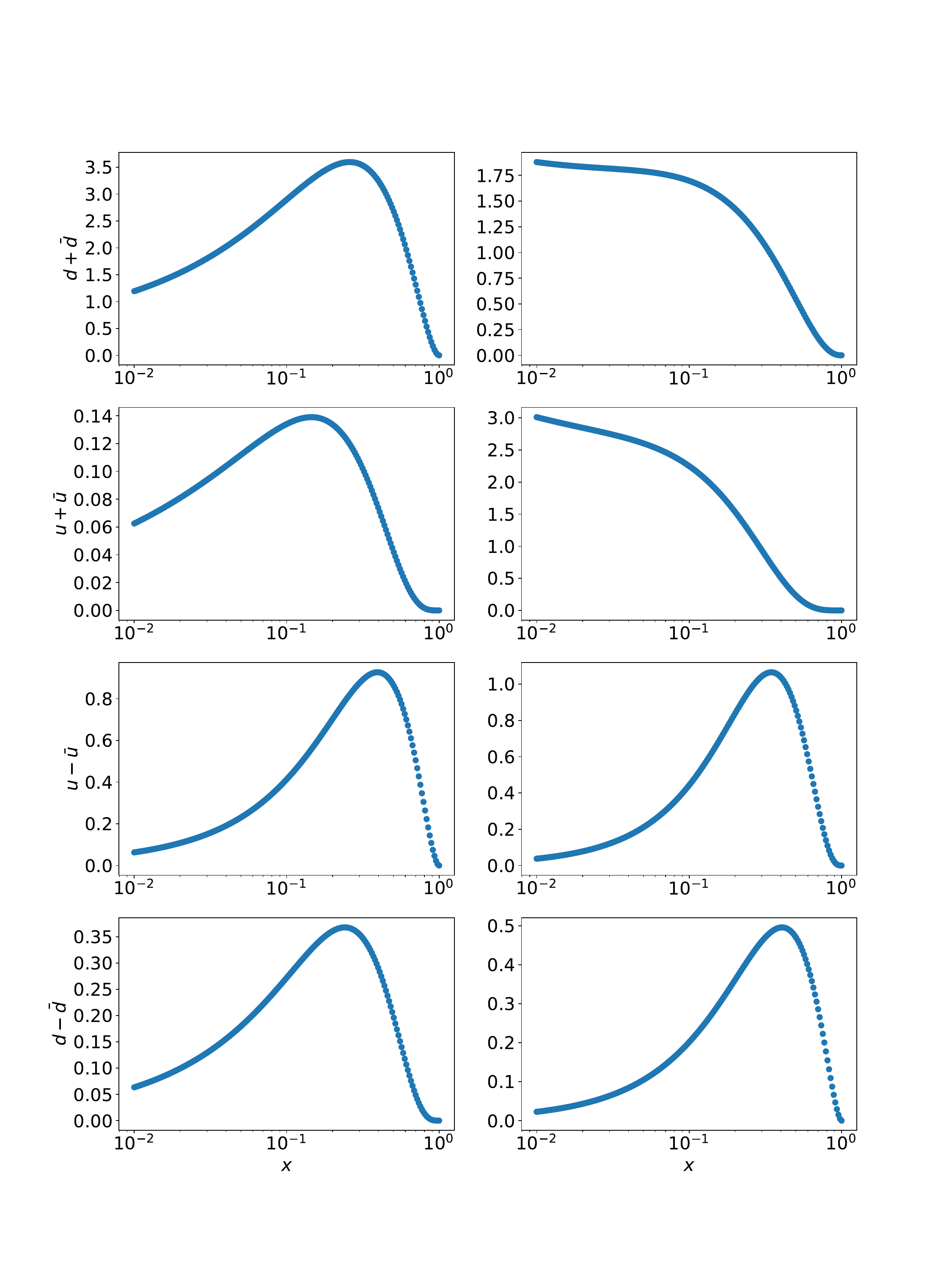}
\vspace*{-0.8cm}
    \caption{(\textit{Left}) Examples of input-layer PDF data for the $q^+$ and $q^-$ combinations with stochastic noise added as inputs into the autoencoder architectures ({\it i.e.}, AE, AE-ML, and AE-WC). The noise corresponds to random fluctuations about the central sampled PDF in each case, where the magnitude of the variations is governed by a Gaussian distribution with a width of
    $10$\% for the $q^+$ distributions and $1$\% for the $q^-$ combinations. (\textit{Right}) Input PDF data examples of $q^+$ and $q^-$ combinations without stochastic noise added; these are used as inputs into the variational autoencoder architectures ({\it i.e.}, VAE and VAIM).}
    \label{fig:pdf_data}
\end{figure*}

Examples taken from the training set in our analysis are shown in Fig.~\ref{fig:pdf_data}; in general, we produce training
data by randomly sampling generic functional forms for each $\mathrm{SU}(2)$ flavor and $C$-even, -odd combination within our
PDF toy model. The initial parametric form, given in Eq.~(\ref{eq:form}) below, has 1 normalization and 4 shape parameters for
each flavor and charge combination to afford a wide range of behaviors in the PDFs' $x$ dependence to train the downstream AE
models. While the generation is random, we fix the seed so that the data sets are consistent across the training and evaluation
of each model. We also introduce a small level of stochastic noise into the data set, especially at large $x$, to simulate
possible uncertainties arising from data as well as provide a regulator to control training as we discuss further below.

We produce $q\! \pm\! \bar{q}$ combinations for the $u$- and $d$-PDFs for a total of 4 independently generated PDF inputs: $d+\bar{d}$, $u+\bar{u}$, $u-\bar{u}$, $d-\bar{d}$ each of which is $x$-weighted when input at the initial encoder layer for training; the 5-parameter form we use for each PDF combination is
\begin{align}
q(x) \pm \bar{q}(x)\, &=\, \mathcal{N}_{q^{\pm}}\, x^{\alpha_{q^{\pm}}}(1 - x)^{\beta_{q^{\pm}}}\, \mathcal{P}_{q^\pm}(x) \nonumber \\
\mathcal{P}_{q^\pm}(x)\, &=\, 1 + \gamma_{q^{\pm}}\sqrt{x} + \delta_{q^{\pm}} x\ ,
\label{eq:form}
\end{align}
in which the adjustable parameters are $\mathcal{N}_{q^{\pm}}$, $\alpha_{q^{\pm}}$, $\beta_{q^{\pm}}$, $\gamma_{q^{\pm}}$, and $\delta_{q^{\pm}}$. Two of these parameters, $\mathcal{N}_{u^{-}}$ and  $\mathcal{N}_{d^{-}}$, are fixed to guarantee consistency with the number sum rules for valence PDFs by integrating the $x$-unweighted $q^-$ PDFs and tuning their normalization constants such that 
$\langle 1 \rangle_{(u^-,\, d^-)} = (2, 1)$, respectively.
Fixing these normalizations thus leaves 18 adjustable parameters for our
toy PDF input model. While this is indeed a toy parametrization, we point out that the dimensional size is roughly comparable to contemporary Hessian global fits, which typically have $\sim\!\!30$ free parameters; as such, the ML framework presented in this study can readily generalize to a comprehensive parametrization of the proton's unpolarized PDFs, including the gluon, strange, and, if phenomenologically justified, fitted charm~\cite{Guzzi:2022rca,Hobbs:2013bia}.

We randomly sample the 18 tunable PDF parameters 10,000 times within a uniform distribution with set limits. Namely, in producing training data, we restrict parameters to
\begin{eqnarray*}
    N_{q^{+}} &\sim& \mathcal{U}(0.25, 1.25) \nonumber \\
    \alpha_{q^{+}} &\sim& \mathcal{U}(-0.25, 0.5)\nonumber \\
    \alpha_{q^{-}} &\sim& \mathcal{U}(0.5, 1.25)\nonumber \\
    \beta_{q^{\pm}} &\sim& \mathcal{U}(1, 5)\nonumber \\
    \gamma_{q^{\pm}} &\sim& \mathcal{U}(0.1, 9.9)\nonumber \\
    \delta_{q^{\pm}} &\sim& \mathcal{U}(0.1, 9.9)\ .
\end{eqnarray*}
We choose a wide range for the PDF parameters in order to limit the prior dependence of our results; the exact prior dependence on the predicted outputs is an open question for most generative models. Numerical values for the $x$-dependent PDFs are sampled logarithmically from $x_{min} = 10^{-2}$ to $x_{max}=0.999$ with 196 points in $x$ per PDF combination. The four PDFs are then concatenated to create an array of 784 $x$-points.
We stress that, beyond this stage, the initial toy-model parametrizations are discarded, such that the various autoencoder
models explored below are agnostic with respect to any underlying functional forms or their associated parametric complexity;
rather, the encoder-decoder networks must dimensionally reduce samplings of 784-dimensional vectors to considerably
more compact latent representations.

As is typical in applications of ML models, we separate the 10,000-member data set into subsets for training, validation, and testing with a 70/15/15-percent split, respectively. The samples are randomly shuffled; however, the random seed is fixed and therefore the same training and validation data sets are used for each model. Similarly, the same test data set is used to assess performance. Finally, we pre-process our data set before it passes through our AE and variational models. The pre-processing procedure we use is standard scaling, which subtracts the mean from each data point and divides by the standard deviation of the full data set.

Given our explicit parametric forms, we calculate Mellin moments for the $q^\pm$ PDFs as shown in Eq.~(\ref{eq:even-odd-moments}) analytically. It should be reiterated that the even moments are constructed from the ``minus" (denoted $q^{-}$)  PDF combinations while the odd moments are constructed from the ``plus" (denoted $q^{+}$) PDFs in agreement with the OPE structure of the lattice-accessible moments; as a result, this physics-based ordering is effectively imposed on our AE models. Because the $q^+$ and $q^-$ PDFs are constructed separately, the moments will occupy two different beta distributions according to the parameters. From the 5-parameter model as specified in Eq.~(\ref{eq:form}), it is possible to compute the corresponding $n^\mathit{th}$ Mellin moment according to:
\begin{widetext}
\begin{eqnarray}
\label{eq:modified_beta}
    \langle x^{n} \rangle_{q} =  \int_{0}^{1} dx\, \mathcal{N}\, x^{n} x^{\alpha}(1-x)^{\beta}(1+\gamma\sqrt{x} + \delta x) && \nonumber \\
    && \hspace{-5cm} = \mathcal{N}\, \Gamma(\beta + 1)\left[\frac{\Gamma(\alpha+n+1)((\alpha + n) \delta + \alpha + n + \beta + \delta + 2)}{\Gamma(\alpha + n + \beta + 3)}\ +\ \frac{\gamma \Gamma(\alpha + n + 3/2)}{\Gamma(\alpha + n + \beta + 5/2)} \right]\ ,
\end{eqnarray}
\end{widetext}
where we have dropped the ``$\pm$'' notation in the expression above, but it is understood that unique moments may be evaluated for the separate $C$-even and -odd PDF combinations.
This analytic solution allows us to generate the moments quickly and independent of any numerical integration method. In our default (and, alternatively, more dimensionally reduced) approach, we evaluate a set of 32 (8) moments in total: 16 (4) moments corresponding to the $u$-PDF and 16 (4) to the $d$-PDF; of these flavor-dependent Mellin moments, 8 (2) are related to the ``+" PDF combinations and 8 (2) are from the complementary ``-" PDFs.
We explore these two alternative dimensionality reductions partly on the basis
of the PCA calculation illustrated in Fig.~\ref{fig:pdf_pca} and related discussion; this showed that a sizable share
of the total PDF variance in our toy model ($\gtrsim 90$\%) can be captured by the 8 leading principle components. We therefore cross-check this finding by significantly restricting our AE and variational models to a much smaller, 8-dimensional latent space. 

For the AE, AE-CL, and AE-WC models, we use training data generated with an assumed stochastic noise, as shown in the two lefthand columns of Fig.~\ref{fig:pdf_data}; meanwhile, for the VAE and VAIM models, we train on input data which do not have this implicit noise (two righthand columns of Fig.~\ref{fig:pdf_data}). In the former scenario, we produce the input noise by taking the central value of the initial sampled PDF and fluctuating about this $x$-dependent central curve according to a Gaussian distribution of assumed width --- typically, $10$\% for the $q^+$ distributions and $1$\% for the $q^-$ combinations. This allows us to introduce some additional randomness into the AE-based models that is naturally present in the VAE models through the randomness of the latent variables.

The stochastic noise effectively functions as a regulator during training of the AE, AE-CL, and AE-WC networks, and we optimize this additional hyperparameter to enhance and ensure convergence.
As with the network topologies themselves, the inclusion of these regulators and similar hyperparameters plays a role in the generalization of a trained ML model for PDF reconstructions and is therefore relevant for stopping criteria that define both the central reconstructed PDFs and their associated uncertainties.
A more comprehensive study of this effect is left to a dedicated paper on uncertainty quantification and comparisons with standard statistical techniques.

\subsection{(Moment) $\to$ (PDF) Reconstructions}
\label{sec:PDF-recon}
We demonstrate the use of encoder-decoder architectures to reconstruct PDFs from input data as well as the encoded moments that are learned in the latent space of each algorithm as compared to the truth moments calculated from the central PDFs. In this section, we focus on the AE-CL and VAIM architectures and what these architectures can achieve once trained. We take these two respective approaches as archetypal of the autoencoder and variational approaches discussed in Sec.~\ref{sec:models}, choosing them in particular as
they satisfy all tractable reconstruction criteria of Tab.~\ref{tab:architectures}.

\begin{figure}[t]
\vspace*{-0.18cm}
\includegraphics[width=0.49\columnwidth]{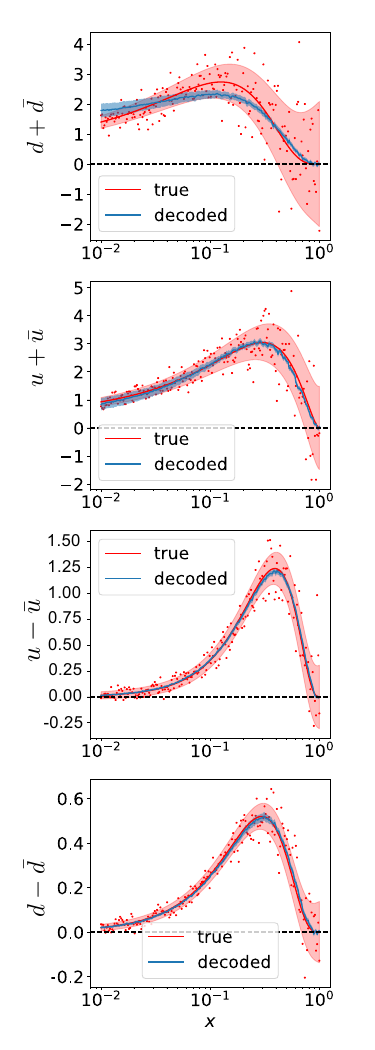}
\hspace*{-0.5cm}\includegraphics[width=0.49\columnwidth]{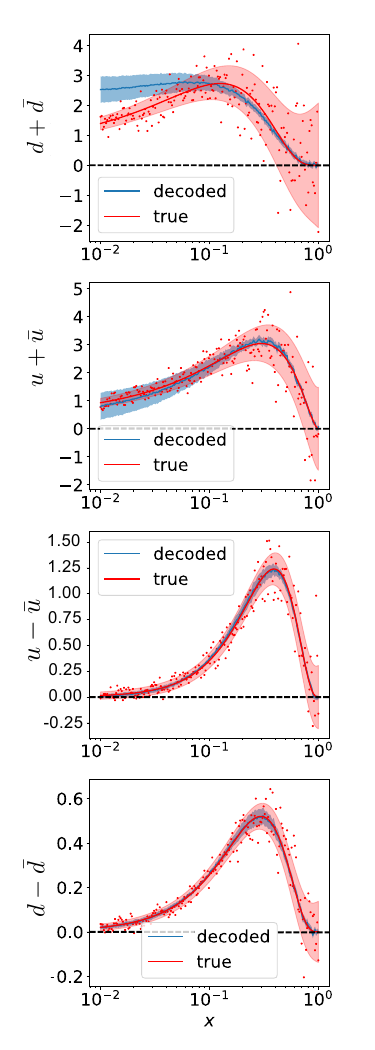}
    \caption{PDF reconstructions from the AE-CL model (blue) as compared to the input data (red points) and the input data distribution (red). The predictions come directly from the encoded PDF moments. (\textit{Left}) The uncertainty in this case is given as the standard deviation over 100 ensemble model predictions. (\textit{Right}) The uncertainty in this case is given as the standard deviation over 100 ensemble model predictions and random sampling from the moment standard deviation.}
    \label{fig:aecl_pdf_reco}
\end{figure}

\subsubsection{AE-CL model PDFs}
In Fig.~\ref{fig:aecl_pdf_reco}, we show the results of the PDF reconstructions from the trained AE-CL algorithm. All plotted results are generated from predictions of the AE-CL model based on the test set; {\it i.e.}, on data which the AE-CL model has not seen during any stage of its training and validation. The solid-red line is the central value of the truth PDF from which the distribution of red points is created. The red points represent random fluctuations about the central truth PDF and are obtained by sampling a Gaussian of defined width as described in Sec.~\ref{sec:data_param}; the shaded red band represents this associated 1$\sigma$ uncertainty on the central value. The blue curves and bands are the predictions of the AE-CL decoder network and are generated by directly sampling the encoded latent space. We show our results for PDF combinations which are $C$-even or -odd, respectively:
\begin{align*}
    &d + \bar{d}\ , \ \ \ \ \ \
    u + \bar{u} \\
    &d - \bar{d}\ , \ \ \ \ \ \
    u-\bar{u}\ .
\end{align*}

\begin{figure}[t]
\vspace*{0.07cm}
\includegraphics[width=\columnwidth]{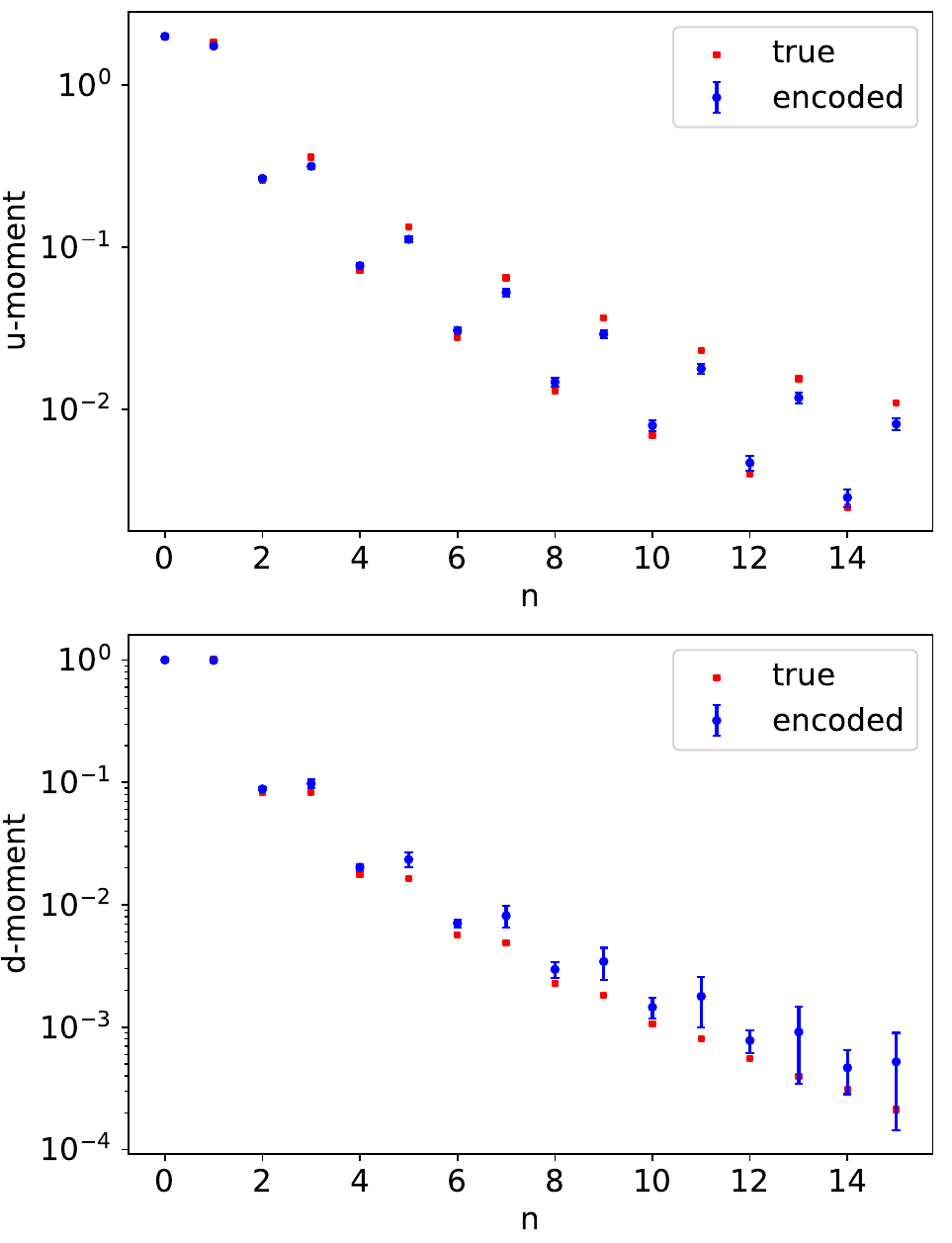}
\vspace*{-0.6cm}
    \caption{Predicted $u$- (top) and $d$-quark (bottom) Mellin moments from the AE-CL encoder (blue) as compared to the truth moments (red). The uncertainty is given by ensembling over a 100-member set of model predictions.}
    \label{fig:aecl_moment_reco}
\end{figure}

We produce decoded results with two approaches to the uncertainty quantification.
The left panels of Fig.~\ref{fig:aecl_pdf_reco} are obtained by ensembling 100 trained AE-CL models, such that the blue line is the mean over the predictions of these 100 models and the blue shaded region is the corresponding 1$\sigma$ uncertainty. The right panels are trained on the same red data set as in the left panels; however, the predictions are generated slightly differently. Instead of sampling the encoded latent directly to create the predictions as in the left panels, we sample Mellin moments within 1$\sigma$ of their associated uncertainties while also ensembling the 100-member collection of model predictions. The righthand panels therefore enfold an additional uncertainty into the central predictions and error bands by randomly varying all accessible model parameters.

This results in a slight shift at low-$x$ of the $d$-quark PDF and a subsequent enlargement of the uncertainties in all cases.
The ability to enfold this additional model uncertainty is a direct consequence of the tractable formulation we have imposed on the (moment) $\to$ (PDF) decoder network, and highlights the value of
building such features into ML models for uncertainty quantification.

The AE-CL decoder predictions reconstruct the truth PDFs with high fidelity, being very close to the original test-set PDFs in Fig.~\ref{fig:aecl_pdf_reco}. There are indications of a mild systematic shift in the $d$-quark PDFs as compared to those of the $u$-quark. This is further reflected in the reconstructed Mellin moments we discuss below, in which the $d$-quark moments seem to be shifted when compared to the truth values, juxtaposed to the $u$-quark moments. We show a small sampling of the full test set which contains 1500 examples. Most of the predictions tend to agree with the test set data within 1$\sigma$ uncertainty; however, there are occasions where there is disagreement. This is not unexpected since there is a significant correlation between the encoded moments in the latent space and the $x$ dependence of the PDFs. Small prediction errors in the Mellin moments can therefore lead to significant deviations of the PDF in $x$-space. A more thorough discussion of these correlated spaces within the encoder-decoder networks is in Sec.~\ref{sec:corr}.


In Fig.~\ref{fig:aecl_moment_reco}, we show the reconstructed $u$- and $d$-quark moments as encoded by the latent space. The red points are the truth values as calculated according to Eq.~(\ref{eq:modified_beta}), while the blue points with uncertainties are from the latent predictions. Since the latent distribution of the AE-CL model is constrained to follow a modified beta distribution, plotting the total array of moments effectively maps the latent space as well. In line with our default AE-CL dimensionality reduction scheme, there are 32 encoded moments, 16 for each flavor. The reconstructed moments follow the general trend of the truth moments; however, there are significant differences among the high $u^+$ moments relative to the truth values unlike the corresponding behavior for the $u^-$ moments. The $d$-quark moments, on the other hand, do mostly agree within their associated uncertainties, although these uncertainties are quite large as compared to the $u$-quark moments.

We quantify uncertainties by sampling the encoded moments and averaging over the predictions of the 100-member model ensemble and evaluating the standard deviation.
Both the $u$- and $d$-quark moments have particularly sizable uncertainties for the $q^{+}$ moments. For the $u$-quark moments the deviations from the truth values are quite large relative to the computed uncertainty, suggesting that the AE-CL model has high confidence despite the inaccuracy of the encoded moments. The $d$-quark moments, on the other hand, are lower in both accuracy and precision, especially for the largest values of $n$. 
This latter behavior may be interpreted in terms of epistemic and aleatoric uncertainties in the sense that
the inherent epistemic uncertainty of the model is reflected in its nominal errors.
Distinguishing between a model's epistemic and aleatoric uncertainties, particularly with respect to
regions of the latent distribution it has not encountered, is a critical point in uncertainty
quantification. We reserve this subject for additional study in the future.

\begin{figure}[t]
\vspace{-0.18cm}
\hspace{-0.5cm}
\includegraphics[width=0.49\columnwidth]{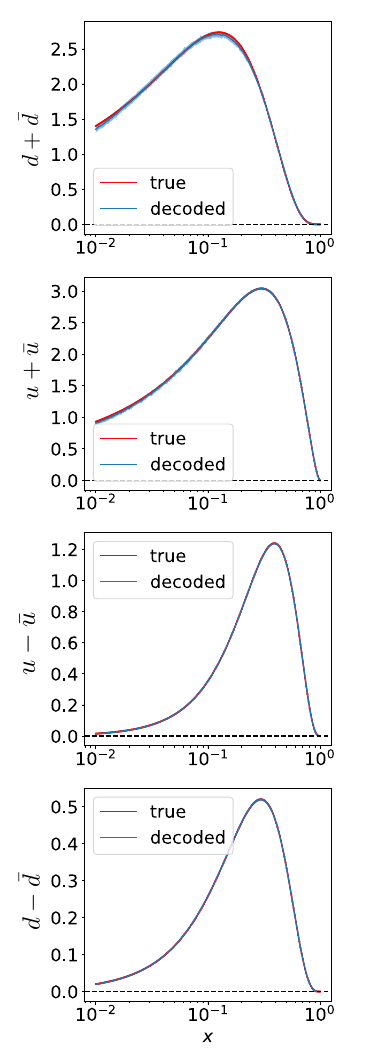}
\hspace*{-0.5cm}\raisebox{-0.1cm}{\includegraphics[width=0.495\columnwidth]{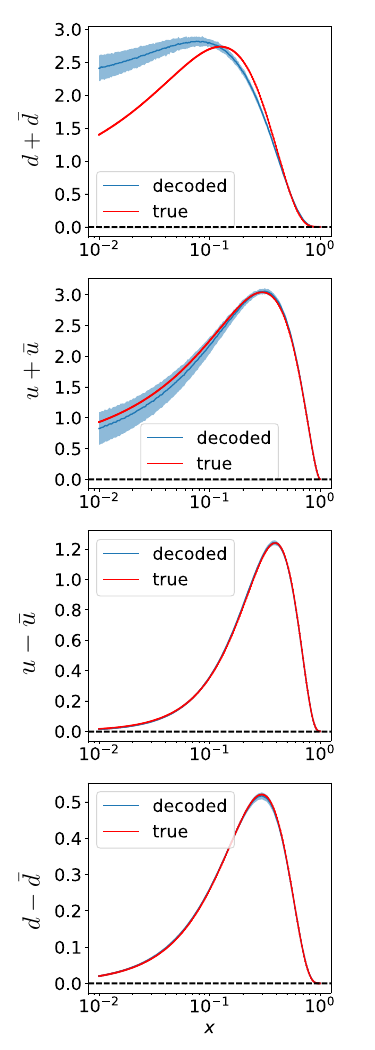}}
\vspace*{-0.1cm}
    \caption{VAIM model reconstructed PDFs (blue) as compared to input PDFs (red). As in Fig.~\ref{fig:aecl_pdf_reco} and described in-text,
    we employ distinct uncertainty prescriptions between the two columns: in the left, the uncertainty is computed from ensembling over 100 model predictions, while, in
    the right panels, we ensemble over random latent variables and network initializations.}
    \label{fig:vaim_pdf_reco}
\end{figure}

\subsubsection{VAIM model PDFs}
Having tested the AE-CL architecture through a detailed exploration of its PDF and moment reconstructions, we turn to the most robust model enumerated in
Tab.~\ref{tab:architectures}, the VAIM approach.
Fig.~\ref{fig:vaim_pdf_reco} shows the results of the VAIM model for PDF reconstructions, randomly plotting two examples from the test set in the left and right
columns, illustrating their separate predictions for $d\!\pm\!\bar{d}$ and $u\!\pm\!\bar{u}$ over the four rows. We show the input test set examples in red for predictions which, for the VAIM calculations, have no added Gaussian noise. 
%
%
To convey the subtleties of error estimation for these network models, we perform the uncertainty quantification
according to distinct prescriptions in each column. In the left panels, the blue curves correspond to the 
means over the 100-member model ensembles while the blue-shaded region is the 1$\sigma$ uncertainty on the associated predictions. For the right panels, however, the uncertainty is given as an ensemble over random latent variables, in the encoded moments uncertainties, and in the initialization of the network model parameters. By systematically varying these parameters around their means and within a standard deviation, we can ensemble the uncertainties and generate marginals of the predictions (which is shown in blue).

\begin{figure}[t]
\vspace*{-0.15cm}
\includegraphics[width=\columnwidth]{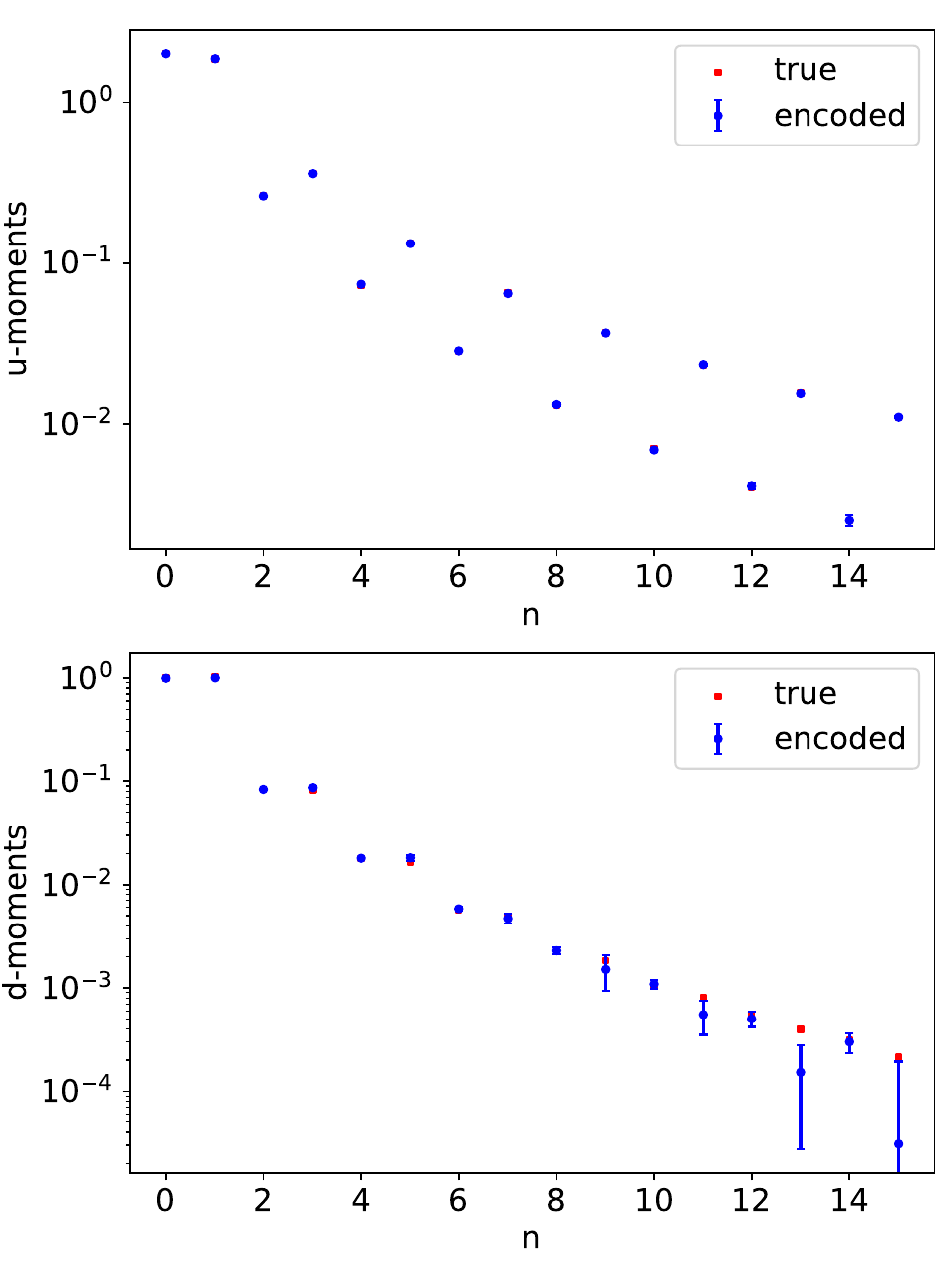}
\vspace*{-0.6cm}
    \caption{Like Fig.~\ref{fig:aecl_moment_reco}, we now plot the VAIM-encoded moments (blue) as compared to the truth values of the corresponding moments (red). The uncertainty is obtained by ensembling over a 100-member set of models predictions. }
    \label{fig:vaim_mom_reco}
\end{figure}

Similarly to the AE-based models, for variational networks there is a shift in the mean when sampling the central value of the moment and varying this quantity as opposed to simply sampling the encoded latent space directly. This leads to a slight deviation from the PDF truth value at low $x$, especially for the $d^+$ PDF reconstruction as seen in the top-right panel of Fig.~\ref{fig:vaim_pdf_reco}.
Ultimately, the decoded uncertainties in both columns are of similar size as those obtained by the AE-based models, while the VAIM predictions seem to sit on the expected results slightly better than for the AE architecture.

As we considered for the AE-CL model, we again examine predictions for the Mellin moments in Fig.~\ref{fig:vaim_mom_reco}, now as obtained via the VAIM model. In the left panel, we plot the first 16 $u$-quark moments; again in accordance with Eq.~(\ref{eq:moment}), the 8 even moments are connected with the underlying $q^-$ PDFs while the 8 odd moments are related to $q^+$. Also as before, we again show the truth-values of the moments as directly computed from the underlying PDFs in red, while the VAIM-predicted, or encoded, values of the Mellin moments are displayed in blue with their predicted uncertainties. Uncertainties and means are generated by ensembling over a 100-member set of model predictions as discussed above for the PDF reconstructions. In the lower panel, we have the same analogous information for the $d$-quark moments. We note that that the $u$-moments are predicted with enhanced accuracy and precision in the VAIM as compared to the AE-CL calculation shown earlier in this subsection; this is similarly true for the $d$-PDF moments 
(including those of the $d^{+}$ combination), despite the fact that, for the largest-order moments, these remain quite uncertain and with larger deviations from the central truth values as seen for
the AE-CL. 

\begin{figure*}[t]
\vspace*{-0.5cm}
    \centering
    \includegraphics[scale=0.18]{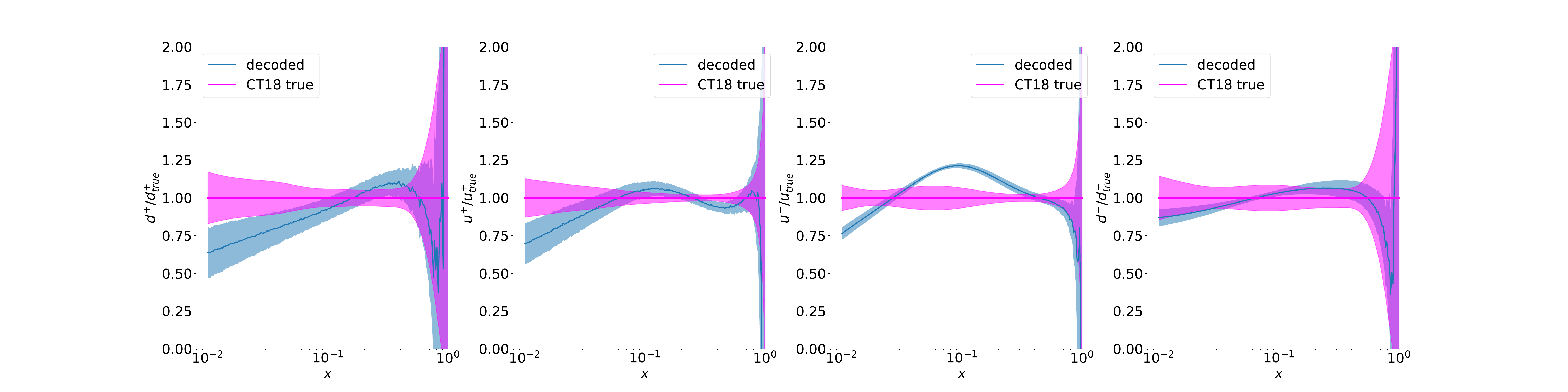} \\
    \vspace*{-0.2cm}\includegraphics[scale=0.18]{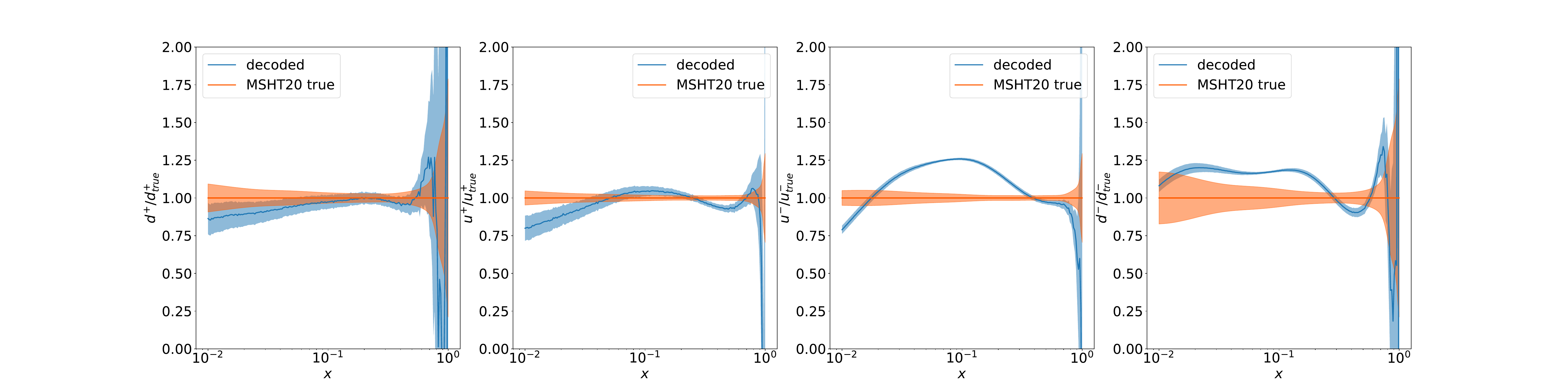} \\
    \vspace*{-0.2cm}\includegraphics[scale=0.18]{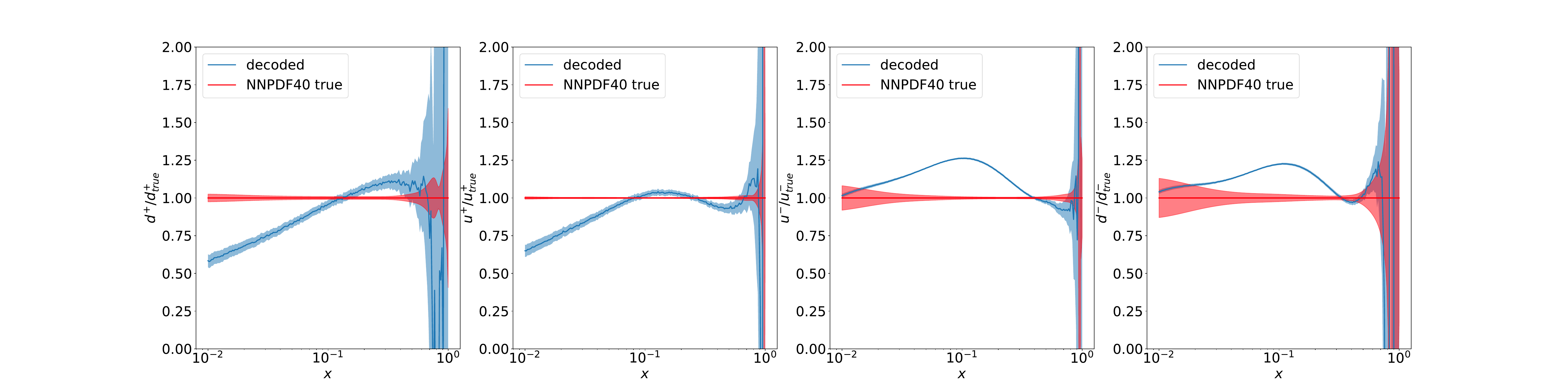}
    \caption{Using the VAIM as a generative model, we compute ratios of predicted PDFs (blue bands) decoded from values of the phenomenological Mellin moments relative to the truth phenomenological PDFs; the moments we input to the decoder are obtained by numerically integrating the phenomenological PDFs. We show direct comparisons for CT18 (top row, magenta bands), MSHT20 (middle row, orange bands), NNPDF4.0 (bottom row, red bands). We supply the decoder with the first 16 $u$- and $d$-quark moments from each respective PDF analysis. To remove effects from scale dependence, we calculate all PDFs at their respective initial scales, $Q_0 = 1.295$, $1.00$, and $1.65$~GeV.}
    \label{fig:ct18nnlo}
\end{figure*}

\subsection{Generating Phenomenological PDFs from the VAIM model}
\label{sec:pheno-PDF}
As a final exploration of the performance of the (moment) $\to$ (PDF) encoder-decoder
architectures, we test an intriguing question:
do the initial-scale parametrizations of the $u$- and $d$-quark PDFs typically used in modern QCD analyses for
HEP
exist as statistically in-distribution solutions of our trained VAIM model?
To explore this possibility concretely, we consider a closely-related question: given an array of Mellin moments for the $u^\pm$ and $d^\pm$
PDFs as specified by a phenomenological fit, how well does the trained VAIM network predict the corresponding fitted PDFs --- {\it i.e.}, with what accuracy
and precision are the phenomenological PDFs predicted by the VAIM given a latent basis of moment values? We afford particularly close attention to the regions of $x$ where the VAIM predictions more closely follow the phenomenological
PDFs on whose moments the encoder-decoder network is trained.

For this exercise, we adopt the most recent baseline fits from three widely-adopted PDF fits used for LHC phenomenology: the NNLO CT18~\cite{Hou:2019qau}, MSHT20~\cite{Bailey:2020ooq}, and NNPDF4.0~\cite{NNPDF:2021njg} PDFs.
For each of these, we calculate the first 16 $\langle x^n \rangle_{u^\pm,\, d^\pm}$ Mellin moments and associated uncertainties while respecting the $C$-even, -odd structure of Eq.~(\ref{eq:even-odd-moments}). As noted above, after directly computing these collections of Mellin moments from public grids, we then take the result
as input to our pre-trained VAIM decoder model.

By construction, our trained VAIM decoder network can be used as a generative model, assuming inputs in the form of the above-mentioned Mellin moments and a random sampling of the latent space --- in this case a uniform distribution. Thus, it is possible to use the decoder sub-network of the VAIM in conjunction with the calculated moments of the NNLO PDFs to test how robustly the VAIM model generates these fits from knowledge of their moments.

The results of this exercise are plotted as the PDF ratios of Fig.~\ref{fig:ct18nnlo}. In all cases, the blue curves represent the ratios of the PDFs predicted by the VAIM model relative to the central phenomenological PDFs, while the associated bands are the VAIM model uncertainties. We directly compare these predictions against the truth PDFs for each phenomenological fit, plotted as magenta (CT), orange (MSHT), and red (NNPDF) curves and bands evaluated at their respective initial scales, $Q_0$. The uncertainties on the VAIM predictions are constructed by oscillating the Mellin moments about their mean values as calculated from the central PDFs using the 1$\sigma$ uncertainty from the error sets of the PDFs.
We note that, in each row, the blue VAIM decoder predictions are based on the respective input moments from each of the phenomenological PDFs against which we compare.

Overall, we note that the PDFs predicted by the VAIM model generally agree within 25\% of the truth PDFs for all $x$ values and flavor combinations; at large $x\! \ge\! 0.1$ --- {\it i.e.}, the region canonically
most correlated with the higher Mellin moments --- the agreement is often quite strong, with the decoded and truth PDFs overlapping within their nominal $1\sigma$ uncertainties. By itself, this result is quite striking, especially considering that the VAIM model has {\it absolutely no information} on the underlying parametrization or theoretical formalism deployed in any of the phenomenological PDFs --- rather, it is merely constrained to agree with respect to the Mellin moments given the decoder. This suggests that significant aspects of the underlying $x$ dependence of the phenomenological PDFs can indeed be encoded by a thoroughly trained variational network, even on the basis of a highly simplified, 5-parameter toy input model. 

Beyond these generation observations, Fig.~\ref{fig:ct18nnlo} also supports several additional
conclusions.
These include the fact that the decoded $u^+$ and $d^+$ PDFs tend to agree to a higher level with
the phenomenological truth PDFs. This agreement is most pronounced for higher $x$ --- especially,
$x \gtrsim 0.1$ --- but can sometimes extend to $x\! \sim\! 10^{-2}$ for the Hessian fits of CT and
MSHT. Nowhere is this robust concurrence more evident than in the decoder model prediction for $d^+(x)$ based on the MSHT
moments, which effectively agrees to high precision with the MSHT20 $d^+$ PDF for all values of $x$ considered
in this study; the analogous comparison with CT18 yields similarly close agreement between the VAIM
prediction and phenomenological PDF, albeit with slightly greater deviations at low $x$.

Meanwhile, the VAIM model also performs quite well in reproducing the $x$ dependence of the $u^+$ PDF, especially for
CT and MSHT, but veers below the NNPDF baseline at low $x$ by $\sim\! 25$\%.
The VAIM predictions for the $C$-odd $q^-$ PDFs tend to deviate more significantly from the
phenomenological PDFs, particularly in the case of the $u^-$ combination plotted in the third column
of Fig.~\ref{fig:ct18nnlo}.
This somewhat greater difficulty in the VAIM performance may reflect challenges in distinguishing
subtle differences in the $x$ dependence of the $q^-$ combination(s) on the basis of the moments.
Even in the scenarios with greatest deviation, however, it is notable that, in the limit of very high
$x\!\gtrsim 0.5$, there is still concordance up to uncertainties between the fitted and decoded
PDFs. The greatest descriptive agreement is achieved for CT18, particularly the $d^-$ combination;
curiously, the decoded $d^-$ PDFs agree with all three phenomenological distributions at
$x\! \sim\! 10^{-2}$, as well as the NNPDF4.0 $u^-$ PDF.
On the other hand, the PDFs for NNPDF and MSHT can disagree at intermediate-to-low values of $x$ by
up to 25\%.

It is also intriguing to compare the relative uncertainties obtained by the encoder-decoder model.
For example, the decoded PDF uncertainty is roughly comparable to that of the CT and MSHT fitted PDFs,
especially for $u^+$ and $d^+$, and, to a lesser extent, for NNPDF for the same combinations. It is
natural to expect some mirroring of PDF errors, since the phenomenological uncertainty on the fitted
PDFs enter the VAIM predictions via the Mellin moments given the generative decoder.
On the other hand, the VAIM PDF uncertainties for the $q^-$ combinations substantially under-predict those of the fitted
PDFs at times, particularly for the NNPDF comparison.
This mixed behavior similarly merits further investigation in the context of uncertainty quantification,
both for fitted PDFs and in ML models as explored here.

Since the latent Mellin space in our default VAIM analysis extends to high order, $n\! \le\! 15$, in reconstructing the $x$ dependence of the PDFs down to $10^{-2}$, we conclude that the autoencoder-based network must introduce some array of internal correlations to leverage its undercompleteness. In the next section, we discuss these correlations further and highlight several points of caution
against too aggressively inferring physics from learned behavior in certain instances.

\subsection{Moment-PDF Correlations for Model Interpretability}
\label{sec:corr}
Lastly, in this section we explore a point noted in the previous discussion: the pattern of correlations between
learned Mellin-space moments and the decoded $x$ dependence of the PDFs. These correlations relate the quantitative
behavior {\it across} the decoder model --- {\it i.e.}, between the learned latent space and reconstructed PDFs --- and 
therefore are the correlations across the generator of the learned likelihood function. Systematic maps of these correlations
can suggest what types of physics information are learned during inferencing in encoder-decoder models for PDFs.

\subsubsection{Internal correlations in neural-network encoder models}
In the AE-CL 
model, the learned physics correlations emerge from mapping the latent distribution to the outputs, whereas in the VAIM model, the analogous correlations develop between the encoded Mellin space, which is distinct from the latent distribution, and the output PDFs. Since these correlations are all across the decoder model, they originate from two different areas of the latent space. For this reason, it is informative to compare correlations between encoded moments and output PDFs, both to understand the stability of these learned correlations under variations in the encoder-decoder architecture and to understand the
physics interpretability of these models.

\begin{figure}
    \centering
    \includegraphics[width=\columnwidth]{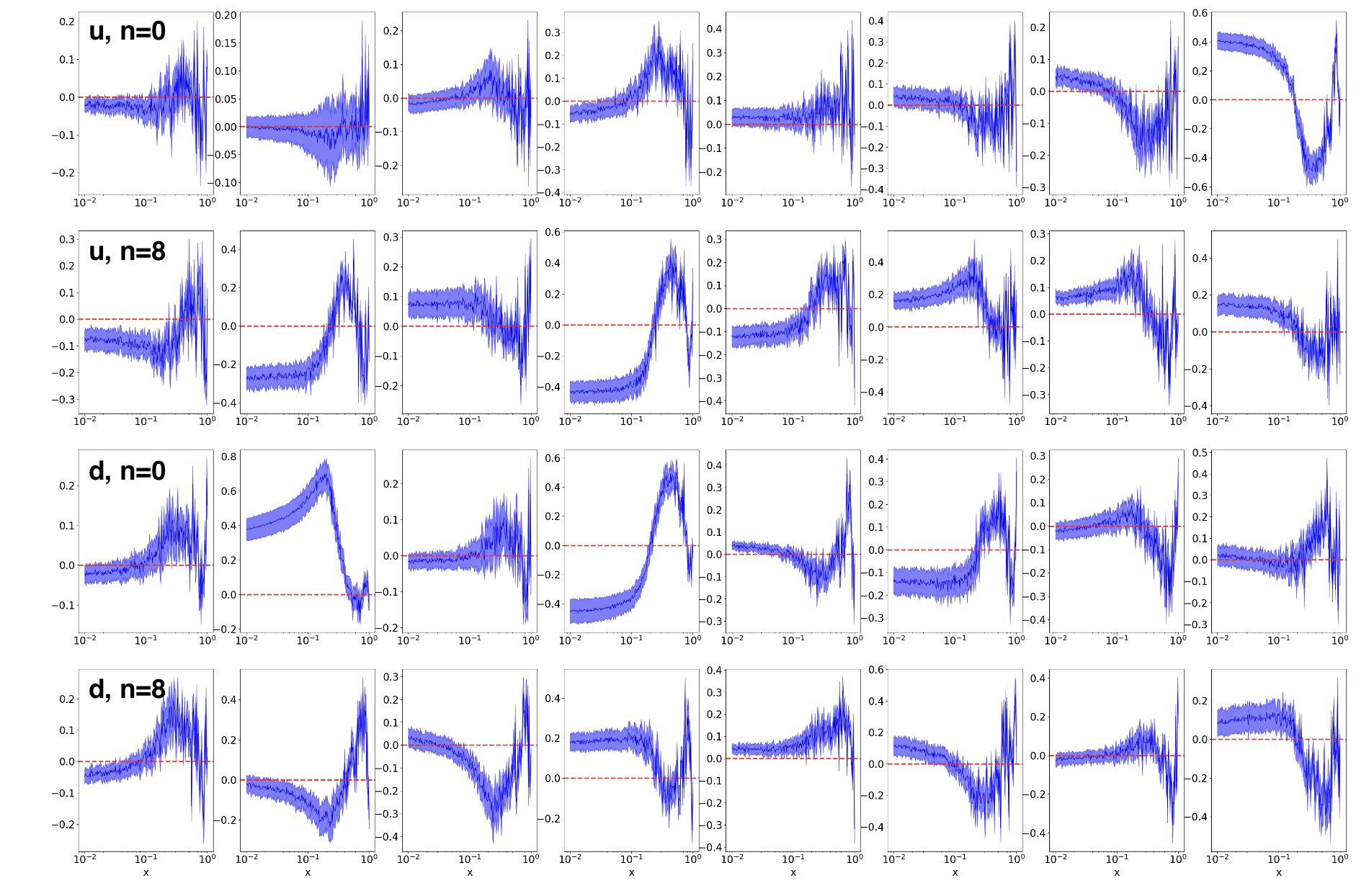}
    \caption{The $x$-dependent correlations between the predicted moments of the $u$-quark (top two rows) and $d$-quark (bottom two rows) and the decoded $d^{+}(x)$ PDF
    ({\it i.e.}, the correlations across the decoder model) as given by the AE-CL architecture. As expected based on the underlying physics, we see the most significant
    correlations in every other panel of the bottom two rows, starting from the second panel from left.}
    \label{fig:aecl_correlations}
\end{figure}

The latent bottleneck inherent to undercomplete autoencoder architectures is crucial to interpreting how these ML-based models learn correlations between the input PDF $x$ dependence and the corresponding Mellin moment. In an undercomplete model, when going from $x$-space to Mellin space, correlated effects expected on the basis of physics knowledge --- {\it e.g.}, between higher-order Mellin moments and high-$x$ PDF dependence --- can become reshuffled. Physics intuition suggests that higher moments should be strongly correlated with the larger-$x$ regions of PDFs since increasing powers of $x$ in $\langle x^n \rangle_{q^\pm}$ effectively suppresses contributions from the low-$x$ behavior of the PDF to the moment. In practice, however, an undercomplete latent bottleneck, acting similarly as PCA, creates many different linear combinations of these distinct $x$ regions during encoding; therefore, there is no reason to expect that the VAIM nor AE-CL models must learn and reflect physical correlations between PDFs and their Mellin-space behavior.

The learning process in encoder-decoder models can thus induce spurious correlations to optimize the goal of
minimal information loss during encoding. Such non-physical encoding effects are most prominently realized in the
model producing correlations between higher Mellin moments and the low-$x$ regions of PDFs minimize loss during reconstruction. The spurious correlations that are learned are not wholly unexpected, as generative models are purposefully constructed to be expressive in their generative power; at the same time, {\it when reading physics into the outputs of these models, one must exercise caution}. Physics interpretability must be built into models systematically and with intention; to that end, tools like the Pearson correlation across different portions of the model can help us interpret the learning process.

\begin{figure}[t]
    \centering
    \includegraphics[width=\columnwidth]{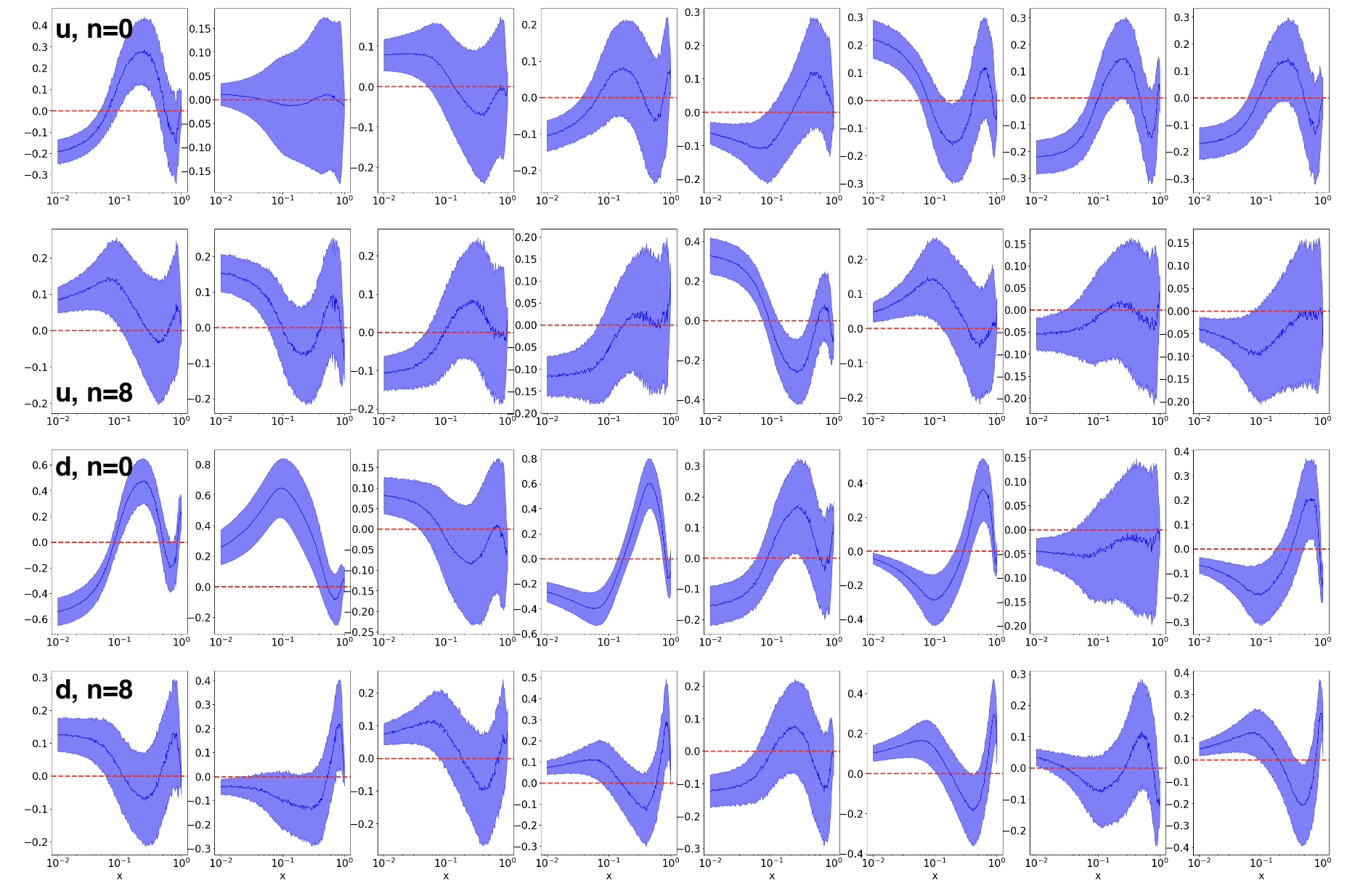}
    \caption{Like Fig.~\ref{fig:aecl_correlations}, but for the analogous correlations obtained within the VAIM decoder
    model.}
\label{fig:vaim_correlations}
\end{figure}

\subsubsection{Calculated moment-PDF correlations}
The ensembling method used above to quantify uncertainties in the encoder-decoder model may be taken as providing a
Monte Carlo sampling of the prediction space; we therefore evaluate correlations according to conventional Monte Carlo
definitions (see, {\it e.g.}, Ref.~\cite{Wang:2018heo}).
The expectation value of a quantity which depends directly on the PDFs can be evaluated over the full ensemble of replicas
as
\begin{eqnarray}
    \langle X \rangle = \frac{1}{N}\sum_{i = 1}^{N} X(q_{i})\ ,
\end{eqnarray}
with the associated Monte Carlo uncertainty given by the standard expression:
\begin{eqnarray}
    \Delta X &=& \sqrt{\frac{1}{N-1}\sum_{i=1}^{N}\Big(X(q_{i}) - \langle X \rangle \Big)^{2}}\ .
\end{eqnarray}
From these quantities, we define the analog of the Pearson correlation for Monte Carlo replicas:
\begin{eqnarray}
\label{eq:corr-def}
\text{Corr}(X,Y) &=& \frac{\langle X Y \rangle - \langle X \rangle \langle Y \rangle}{\Delta X \Delta Y}\ ;
\end{eqnarray}
using this latter expression in Eq.~(\ref{eq:corr-def}), we compute correlations between sectors of the
encoder-decoder networks from our set of ensembled replicas.

Starting first with the autoencoder architecture, in Fig.~\ref{fig:aecl_correlations}, we plot correlations in
the trained AE-CL decoder model between the latent distribution of Mellin moments and the output PDFs; for the sake
of this discussion, we compute correlations between the full array of latent Mellin moments and the $d^{+}$ PDF. We take
this set of correlations as an example, but note that qualitatively similar patterns are obtained for the
analogous correlations with other PDF flavor and charge combinations.

For the AE-CL model, the latent distribution is directly constrained to the Mellin moments during training. In this figure, the two upper rows plot the $x$-dependent correlations between the various $u$-quark Mellin moments, $\langle x^n \rangle_{u^\pm}$, and the $d^{+}(x)$ PDF; the bottom two rows give the $x$-dependent correlations between the $d$-quark Mellin moments, $\langle x^n \rangle_{d^\pm}$, and $d^{+}(x)$. The panels are organized such that the top left panel of the $u$- and $d$-PDF moments correspond to $n\!=\!0$, which is calculated using the $u^{-}$ and $d^{-}$ PDFs. The next panel then corresponds to the correlation with the $n\!=\!1$ moment, which is calculated using the $u^{+}$ and $d^{+}$ PDFs. Consistent with Eq.~(\ref{eq:even-odd-moments}), the correlations in the panels alternate in this manner for the complete set of 16 moments, yielding a total of 32 moment-PDF correlation. As before, uncertainties are from ensembling the correlations calculated from sampling moments within their encoding uncertainty and statistically combining across all models in the ensemble. The uncertainty bands capture global properties of these correlations by eliminating possible statistical fluctuations in an individual model.

\begin{figure}[t]
    \centering
    \includegraphics[width=\columnwidth]{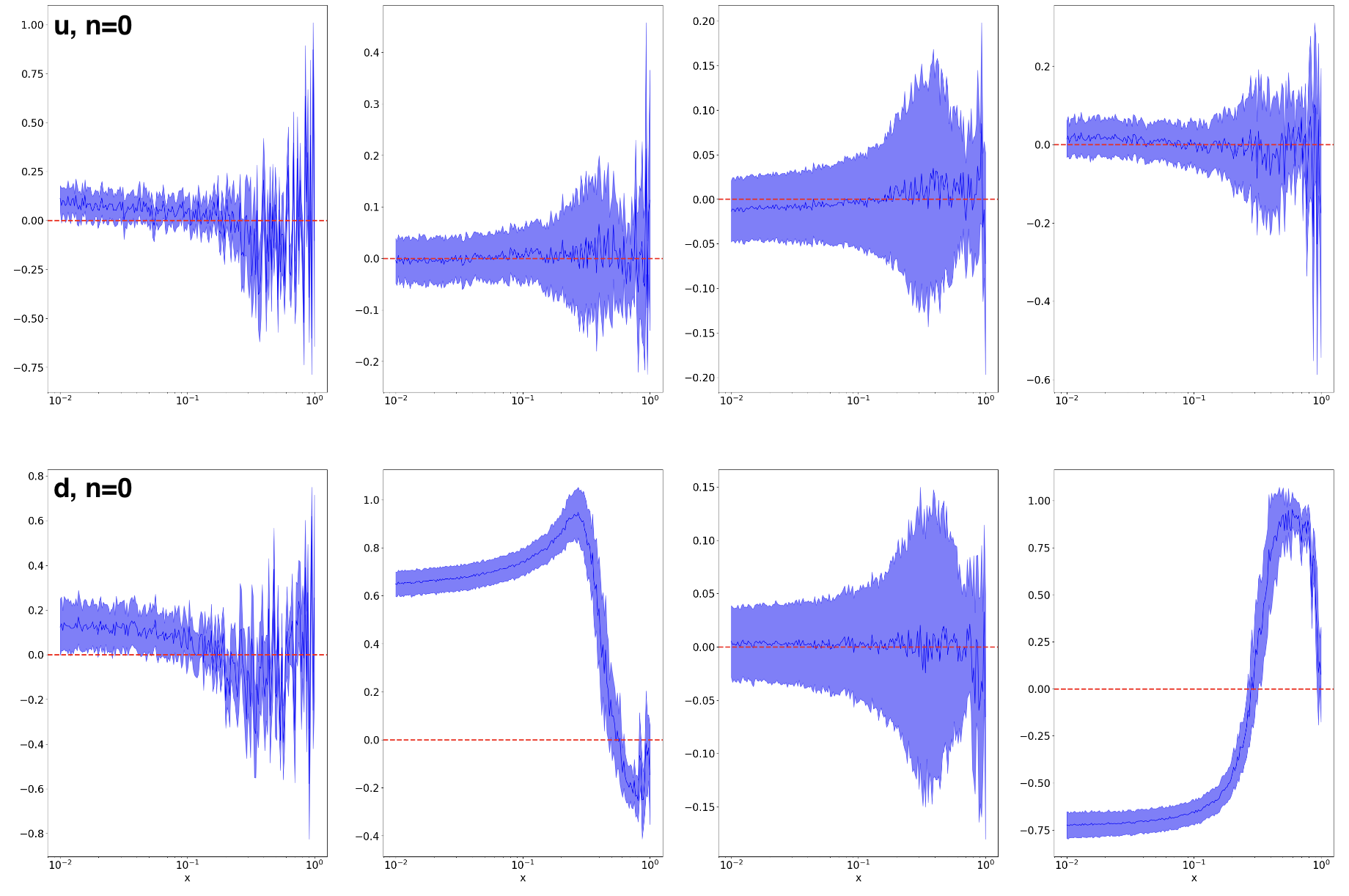}
    \includegraphics[width=\columnwidth]{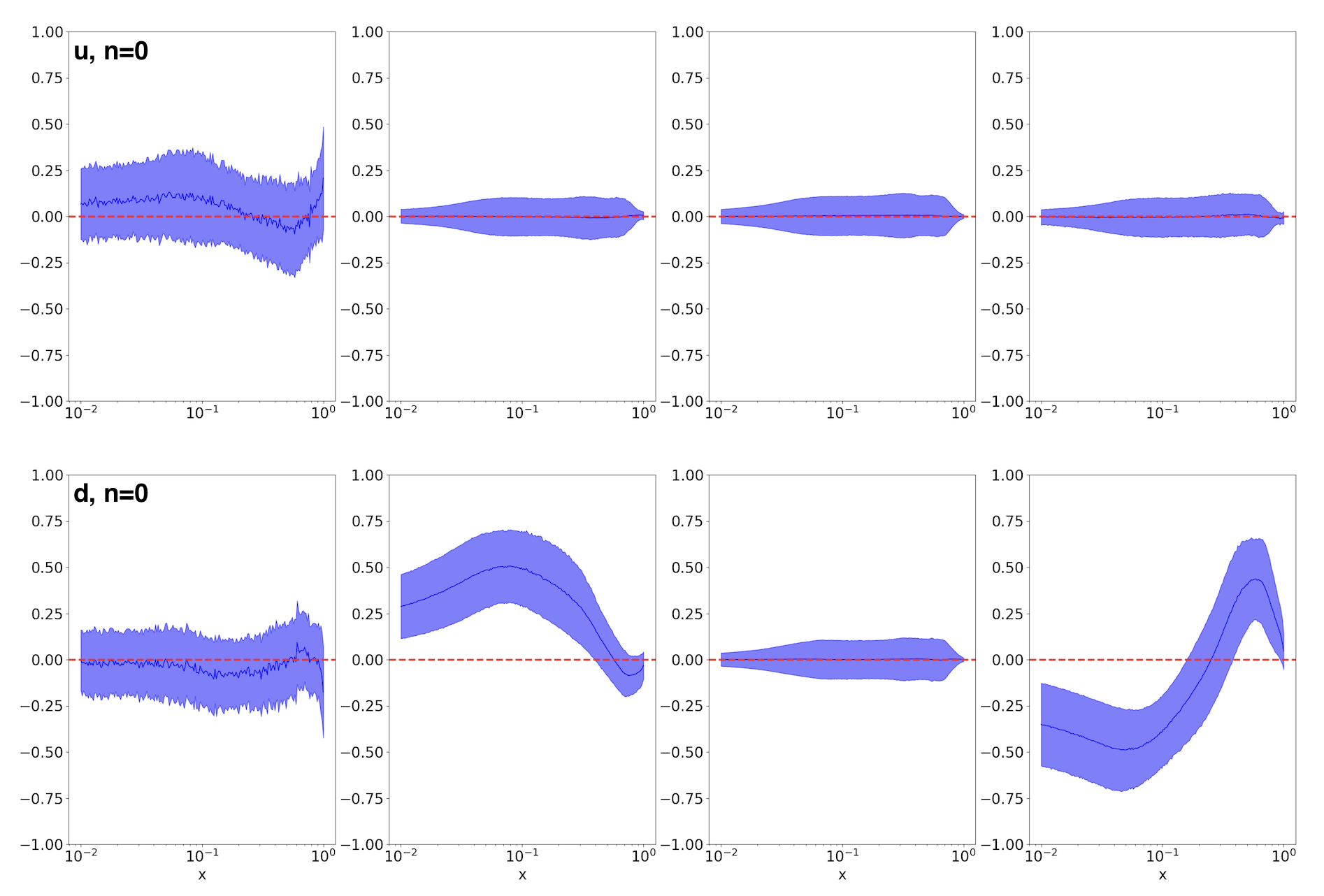}
    \caption{As in Figs.~\ref{fig:aecl_correlations} and~\ref{fig:vaim_correlations}, but now
    plotting correlations in the more dimensionally-reduced scenario, {\it i.e.}, between 
    only eight encoded moments and the decoded $d^{+}(x)$ PDF as
    obtained by the AE-CL model (upper two rows) and VAIM (lower two rows).}
    \label{fig:dim8_correlations}
\end{figure}

We would expect strong correlations between the $d^{+}$ PDF and the moments calculated from the $d^{+}$ PDF, and in particular for the correlations to peak at successively larger values of $x$ for moments of higher order. This expected behavior is indeed realized in the AE-CL model, reflected in significant correlations between the $d^+$ PDF and the various $\langle x^n \rangle_{d^+}$ moments. These correlations generally tend to peak at very high $x$, especially for increasing choices of the moment order, $n$. Many of the other correlations are zero within uncertainties, which agrees with our intuition. At the same time, a small number of sizable correlations between the $d^{+}$ PDF and the $u^{\pm}$ and $d^{-}$ moments are generated as well; these correlations can span the full $x$ range from $x = 10^{-2}$ to $x = 1$. 

Similarly, we see in Fig.~\ref{fig:vaim_correlations} the $x$-dependent correlations between the $d^{+}$ PDF and the $u^{\pm}$ and $d^{\pm}$ moments across the trained VAIM decoder where the panels are organized as they are in Fig.~\ref{fig:aecl_correlations}. One can see that the correlations have larger uncertainty due to the increased randomness of the variational autoencoder model. The correlations between the $d^{\pm}$ moments and the $d^{+}$ PDF seem to have similar shapes between the VAIM and AE-CL models, unlike the correlations between the $u^{\pm}$ moments and the $d^{+}$ PDF, which behave in a discrepant manner between the two models. This
basic pattern is consistent with expectation, since the Mellin structure imposed on the latent ensures preservation of physical
correlations between the moments and their corresponding PDFs; outside this constraint, the architectures we explore are free to
optimize in a fashion that does not introduce a systematic set of correlations among otherwise independent flavor and charge
combinations.

As discussed in Sec.~\ref{sec:data_param}, another crucial degree-of-freedom in configuring the latent space is the number of dimensions reserved for encoding. Our default for the calculations in this study has been to encode in 32 latent dimensions. According to the dimensionality reduction calculation in Sec.~\ref{sec:dimred}, 32 dimensions is more than adequate to encode maximal information with minimal loss. It is interesting, however, to test the sufficiency of a more aggressive dimensional reduction, for
example, to explore whether the pattern of correlations observed earlier between the PDFs and lower moments survives this greater compression.

To this end, we also train a series of still more undercomplete networks for the AE-CL and VAIM models, taking a latent distribution of only 8 dimensions. This ensures that there are 8 total moments encoded, thereby limiting the available domain in the latent space  for the encoder model to reshuffle information on the $x$ dependence of the PDFs. We show in Fig.~\ref{fig:dim8_correlations} the correlations between the $u^{\pm}$ and $d^{\pm}$ moments and the $d^{+}$ PDF with an 8-dimensional latent representation. Notice that, in the case of the lower-dimensional latent space, the large correlations between the $d^{+}$ PDF and the $u^{\pm}$ and $d^{-}$ moments have now stabilized to 0 for the entire range of $x$ and that the correlations only appear to be with the $d^{+}$ moments.

This observed behavior suggests that further reductions in the dimensionality of the latent vector preserve dominant
physical correlations with decoded PDFs while suppressing the tendency of the variational
model to encode $x$ dependence over a wider array of extraneous dimensions. Although this reduced latent
may modestly circumscribe the generative power of the VAIM model, it nonetheless further enhances
the interpretability of the latent space. This greater interpretability is observable in the vanishing
of spurious correlations between moments and decoded PDFs of different flavor.

\section{Conclusions}
\label{sec:conclusions}

In this study, we have for the first time explored a tractable encoder-decoder framework, \texttt{PDFdecoder},
to represent PDFs in terms of their Mellin space behavior.
We have taken Mellin moments as a demonstration case given the rapid growth of lattice methods
to refine knowledge of PDFs, especially at high $x$; this subject also attracts significant phenomenological
interest as relative uncertainties on the PDFs at $x\! \ge\! 0.1$ limit the precision of searches for BSM
physics.
The \texttt{PDFdecoder} approach is of significant flexibility and generality that it can be used to understand
the properties of many PDF sets and models, with the potential for various extensions; for the sake of this analysis, we
restricted ourselves to simplified toy models of the PDFs in the flavor-$\mathrm{SU}(2)$ sector.
More generally, this provides a context to explore issues related to the parametrization
of PDFs using neural networks based on multi-layer perceptrons, which provide the core
structure of the encoder-decoder models trained in this work.

By framing our study in terms of encoder-decoder networks, we were able to investigate
the systematics of these models by varying network architectures, configurational aspects of
latent vectors, and associated hyperparameters.
As a consequence, we identified a number of essential aspects necessary to improve ML
models in the direction of greater interpretability.
During the course of this analysis, we developed a series of criteria to ensure tractability
and interpretability in reconstructing PDFs from learned behavior in Mellin space, and ultimately
identified two main architectures which most robustly satisfy these criteria: the autoencoder with
constrained latent (AE-CL model) and variational autoencoder inverse mapper (VAIM model).

Having identified these specific network topologies, we systematically tested their performance as
generative models to decode the PDFs from information on their Mellin moments. We performed this task as
a demonstration study, first assuming extensive knowledge of the Mellin behavior in 32-dimensional
latent spaces; on this basis, we found very robust performance in PDF reconstruction, suggesting the possibility
of decoding such models from practical information as might be furnished by lattice-gauge calculations or
other empirical sources.

Ultimately, we found the VAIM model to provide the greatest combination in generative capacity as well as interpretability
in the (moment) $\to$ (PDF) decoding problem. As a demonstration, we used a pre-trained VAIM model to predict the $x$-dependent behavior
of select phenomenological PDFs from knowledge of their Mellin moments. This exercise resulted in an intriguing
pattern of agreement and deviation, with the encoder-decoder network succeeding in reproducing fitted PDFs
at high $x$ within uncertainties, and often performing well even at quite low $x$, particularly for
$C$-even PDF combinations, $q^+(x)$.

We further investigated the (moment) $\to$ (PDF) decoding by computing cross-network correlations, thereby
introducing a novel tool to interpret the content in encoded latent spaces.
With this tool, we were able to explore the differing performances of unique encodings, including the
reduction of the latent space to smaller dimensionality.
In this case, we find a trade-off between the fidelity of the PDF reconstruction and the interpretability
of the latent as the size of the latter is increased.
Collectively, these findings imply a novel line of research in developing PDF parametrizations and
interpreting neural-network-based representations of PDFs.

Multiple extensions of this demonstration study naturally suggest themselves, including the exploration of more comprehensive parametrizations of the
full flavor dependence of the PDFs, as well as related objects in HEP phenomenology.
In this context, additional theoretical ingredients might also be incorporated and potentially disentangled, especially with respect to
their immediate impact upon the fitted PDFs.
We have also extended the use of correlation-derived metrics to encoder networks as an interpretability tool to reveal associations that
neural networks can learn when making predictions; this might be extended in further studies of the issues above and in tests of the
durability of learned correlations under additional network assumptions and constraints.

Further exploring uncertainty quantification in this framework is necessary. This would entail more exhaustive testing of the assumptions
in the sampling and ensembling methods used here as well as consideration of other approaches like
incorporation of Bayesian neural networks, approximations with Dropout, and Dirichlet prior networks.
In this setting, investigating interpretations of PDF uncertainties derived from Hessian-based tolerance methods and Monte Carlo sampling
could be another vital application of the \texttt{PDFdecoder} approach.

We make a number of core modules of the \texttt{PDFdecoder} framework public in an associated GitHub
repository, \url{https://github.com/ML4HEP-Theory/PDFDecoder}. This code will allow the construction and training of the AE-CL and VAIM models,
while full data sets and pre-trained models leading to the main results of this analysis 
can be made available by the authors upon request.

\begin{acknowledgements}
We are thankful to colleagues in the Argonne HEP Division and CELS Directorate for stimulating exchanges related to the subject of this calculation.
TJH thanks members of the CTEQ-TEA Collaboration for helpful discussions on PDF uncertainties and parametrizations.
This work at Argonne National Laboratory was supported by the U.S.~Department of Energy under contract DE-AC02-06CH11357.

\end{acknowledgements}

\appendix

\section{Principal Component Analysis}
\label{sec:pca}

Principal component analysis (PCA) is often considered an unsupervised dimensionality reduction technique on the grounds that
the transformations involved are done purely on the unlabeled feature space of the data. In fact, PCA can be derived from basic
linear algebra principles as we outline here.
We present this brief mathematical argument as a supplement to the discussion in Sec.~\ref{sec:dimred}, the point being the
direct relation between simplified, linear autoencoder architectures and PCA. 

We consider a data set with a total number of features, $n$, such that we wish to encode this data into a reduced dimensional representation of itself of size $l$ with as little loss as possible. A single example of this data is given as a column vector, $x \in \mathbb{R}^{n}$. In this case, we project to a coding space, $z \in \mathbb{R}^{l}$. The decoder can be constructed by matrix multiplication, $g(z) = Dz$, where $D \in \mathbb{R}^{n\times l}$ such that $g(z) \in \mathbb{R}^{n}$. The tensor $D$ is pseudo-orthonormal because we impose the condition that the columns of $D$ are orthogonal. This condition means that $D^{T}D = \mathbb{1}^{l\times l}$ whereas $DD^{T} \neq \mathbb{1}^{n\times n}$

We can use the squared Euclidean distance (or Frobenius norm) to find the optimal codes, $z^{*}$, such that the decoded data have minimal loss relative to the input data. The goal is develop an encoder such that in the reduced representation latent space, there is maximal preservation of structure,
\begin{eqnarray}
   z^{*} = \argmin_{z}{\|x - g(z) \|^{2}_{2}}\ .
\end{eqnarray}
We can expand this expression using the transpose operator, and collect like terms (relying, for instance, on the fact that $x^{T}g(z) = g^{T}(z)x$, given that this quantity is a scalar) to find:
\begin{eqnarray}
    z^{*} &=& \argmin_{z}\left(-2x^{T}g(z) + g^{T}(z)g(z)\right) \nonumber \\
    &=&  \argmin_{z}\left(-2x^{T}Dz + z^{T}z\right)\ ,
\end{eqnarray}
where, in the last line, we substituted the tensor-product definition of $g(z)$ and used the identity, $D^{T}D = \mathbb{1}$.

We can find the optimal value of the codes by setting the gradient with respect to $z$ equal to 0, and solving for $z$:
\begin{eqnarray}
    0 = \nabla_{z}\left(-2x^{T}Dz + z^{T}z \right)\ ;
\end{eqnarray}
this latter expression implies
\begin{eqnarray}
    z = D^{T}x\ .
\end{eqnarray}
We see that the encoding function can be expressed as $e(x) = z = D^{T}x$, and the decoding function as $g(z) = Dz$, meaning that the output, $x'$, of the full autoencoding process is given as
\begin{eqnarray}
    x^{\prime} = g\big[ e(x) \big] = DD^{T}x\ .
\end{eqnarray}
Similarly as before, we again use the squared Euclidean distance to measure the distance between the outputs of the PCA and the inputs to find the optimal encoding tensor $D^{*}$,
\begin{eqnarray}
    D^{*} = \argmin_{D}\sum_{i}\| x_{i} -  DD^{T}x_{i} \|_{2}^{2}\ ,
\end{eqnarray}
where $i$ is the label of example $x_{i}$ as part of a larger data set, $x_{i} \in X$. Now consider that $x_{i}^{T}$ is a row in $X$, where $i$ runs from $1$ to $m$. Then the full data set feature space is given by $X \in \mathbb{R}^{m\times n}$. We can then re-express the previous equation by eliminating the explicit summation over $i$,
\begin{eqnarray}
    D^{*} = \argmin_{D}\| X -  XDD^{T} \|_{2}^{2}\ .
\end{eqnarray}
By using the relation between the Frobenius (Euclidean) norm and the trace, $\| A\|_{2}^{2} = Tr \left(A A^{T} \right)$ while expanding the resulting expression and eliminating terms, the optimal encoder is therefore given by
\begin{eqnarray}
    D^{*} &=& \argmin_{D}\left( - D^{T}X^{T}X D\right) \nonumber \\
    &=& \argmax_{D}\left( D^{T}X^{T}X D\right)\ .
\label{eq:Dvar}
\end{eqnarray}

If the data matrix above, $X$, is shifted by the mean of each column, then Eq.~(\ref{eq:Dvar}) becomes a statement respecting the covariance matrix; that is, the optimal encoding matrix maximizes the variance of the data feature space while simultaneously minimizing the loss between the input and the subsequent decoded values.

\bibliographystyle{utphys}
\bibliography{mellin}

\providecommand{\noopsort}[1]{}\providecommand{\singleletter}[1]{#1}%
\providecommand{\href}[2]{#2}\begingroup\raggedright\begin{thebibliography}{10}

\bibitem{Apollinari:2017lan}
``{High-Luminosity Large Hadron Collider (HL-LHC)}: {Technical Design Report V.
  0.1},'' 2017.

\bibitem{Amoroso:2022eow}
S.~Amoroso {\em et~al.}, ``{Snowmass 2021 Whitepaper: Proton Structure at the
  Precision Frontier},''
  \href{http://dx.doi.org/10.5506/APhysPolB.53.12-A1}{{\em Acta Phys. Polon. B}
  {\bfseries 53} no.~12, (2022) 12--A1},
  \href{http://arxiv.org/abs/2203.13923}{{\ttfamily arXiv:2203.13923
  [hep-ph]}}.

\bibitem{ZEUS:2019cou}
{\bfseries ZEUS} Collaboration, H.~Abramowicz {\em et~al.}, ``{Limits on
  contact interactions and leptoquarks at HERA},''
  \href{http://dx.doi.org/10.1103/PhysRevD.99.092006}{{\em Phys. Rev. D}
  {\bfseries 99} no.~9, (2019) 092006},
  \href{http://arxiv.org/abs/1902.03048}{{\ttfamily arXiv:1902.03048
  [hep-ex]}}.

\bibitem{Carrazza:2019sec}
S.~Carrazza, C.~Degrande, S.~Iranipour, J.~Rojo, and M.~Ubiali, ``{Can New
  Physics hide inside the proton?},''
  \href{http://dx.doi.org/10.1103/PhysRevLett.123.132001}{{\em Phys. Rev.
  Lett.} {\bfseries 123} no.~13, (2019) 132001},
  \href{http://arxiv.org/abs/1905.05215}{{\ttfamily arXiv:1905.05215
  [hep-ph]}}.

\bibitem{Greljo:2021kvv}
A.~Greljo, S.~Iranipour, Z.~Kassabov, M.~Madigan, J.~Moore, J.~Rojo, M.~Ubiali,
  and C.~Voisey, ``{Parton distributions in the SMEFT from high-energy
  Drell-Yan tails},'' \href{http://dx.doi.org/10.1007/JHEP07(2021)122}{{\em
  JHEP} {\bfseries 07} (2021) 122},
  \href{http://arxiv.org/abs/2104.02723}{{\ttfamily arXiv:2104.02723
  [hep-ph]}}.

\bibitem{Madigan:2021uho}
M.~Madigan and J.~Moore, ``{Parton Distributions in the SMEFT from high-energy
  Drell-Yan tails},'' \href{http://dx.doi.org/10.22323/1.398.0424}{{\em PoS}
  {\bfseries EPS-HEP2021} (2022) 424},
  \href{http://arxiv.org/abs/2110.13204}{{\ttfamily arXiv:2110.13204
  [hep-ph]}}.

\bibitem{Gao:2022srd}
J.~Gao, M.~Gao, T.~J. Hobbs, D.~Liu, and X.~Shen, ``{Simultaneous CTEQ-TEA
  extraction of PDFs and SMEFT parameters from jet and $ t\overline{t} $
  data},'' \href{http://dx.doi.org/10.1007/JHEP05(2023)003}{{\em JHEP}
  {\bfseries 05} (2023) 003}, \href{http://arxiv.org/abs/2211.01094}{{\ttfamily
  arXiv:2211.01094 [hep-ph]}}.

\bibitem{CMS:2021yzl}
{\bfseries CMS} Collaboration, A.~Tumasyan {\em et~al.}, ``{Measurement and QCD
  analysis of double-differential inclusive jet cross sections in proton-proton
  collisions at $ \sqrt{s} $ = 13 TeV},''
  \href{http://dx.doi.org/10.1007/JHEP02(2022)142}{{\em JHEP} {\bfseries 02}
  (2022) 142}, \href{http://arxiv.org/abs/2111.10431}{{\ttfamily
  arXiv:2111.10431 [hep-ex]}}. [Addendum: JHEP 12, 035 (2022)].

\bibitem{Iranipour:2022iak}
S.~Iranipour and M.~Ubiali, ``{A new generation of simultaneous fits to LHC
  data using deep learning},''
  \href{http://dx.doi.org/10.1007/JHEP05(2022)032}{{\em JHEP} {\bfseries 05}
  (2022) 032}, \href{http://arxiv.org/abs/2201.07240}{{\ttfamily
  arXiv:2201.07240 [hep-ph]}}.

\bibitem{Fu:2023rrs}
Y.~Fu, R.~Brock, D.~Hayden, and C.-P. Yuan, ``{Probing Parton distribution
  functions at large x via Drell-Yan Forward-Backward Asymmetry},''
  \href{http://arxiv.org/abs/2307.07839}{{\ttfamily arXiv:2307.07839
  [hep-ph]}}.

\bibitem{Ablat:2023tiy}
A.~Ablat, M.~Guzzi, K.~Xie, S.~Dulat, T.-J. Hou, I.~Sitiwaldi, and C.~P. Yuan,
  ``{Exploring the impact of high-precision top-quark pair production data on
  the structure of the proton at the LHC},''
  \href{http://arxiv.org/abs/2307.11153}{{\ttfamily arXiv:2307.11153
  [hep-ph]}}.

\bibitem{Jing:2023isu}
X.~Jing {\em et~al.}, ``{Quantifying the interplay of experimental constraints
  in analyses of parton distributions},''
  \href{http://dx.doi.org/10.1103/PhysRevD.108.034029}{{\em Phys. Rev. D}
  {\bfseries 108} no.~3, (2023) 034029},
  \href{http://arxiv.org/abs/2306.03918}{{\ttfamily arXiv:2306.03918
  [hep-ph]}}.

\bibitem{Courtoy:2020fex}
A.~Courtoy and P.~M. Nadolsky, ``{Testing momentum dependence of the
  nonperturbative hadron structure in a global QCD analysis},''
  \href{http://dx.doi.org/10.1103/PhysRevD.103.054029}{{\em Phys. Rev. D}
  {\bfseries 103} no.~5, (2021) 054029},
  \href{http://arxiv.org/abs/2011.10078}{{\ttfamily arXiv:2011.10078
  [hep-ph]}}.

\bibitem{Hobbs:2019gob}
T.~J. Hobbs, B.-T. Wang, P.~M. Nadolsky, and F.~I. Olness, ``{Charting the
  coming synergy between lattice QCD and high-energy phenomenology},''
  \href{http://dx.doi.org/10.1103/PhysRevD.100.094040}{{\em Phys. Rev. D}
  {\bfseries 100} no.~9, (2019) 094040},
  \href{http://arxiv.org/abs/1904.00022}{{\ttfamily arXiv:1904.00022
  [hep-ph]}}.

\bibitem{Wang:2018heo}
B.-T. Wang, T.~J. Hobbs, S.~Doyle, J.~Gao, T.-J. Hou, P.~M. Nadolsky, and F.~I.
  Olness, ``{Mapping the sensitivity of hadronic experiments to nucleon
  structure},'' \href{http://dx.doi.org/10.1103/PhysRevD.98.094030}{{\em Phys.
  Rev. D} {\bfseries 98} no.~9, (2018) 094030},
  \href{http://arxiv.org/abs/1803.02777}{{\ttfamily arXiv:1803.02777
  [hep-ph]}}.

\bibitem{Kotz:2023pbu}
L.~Kotz, A.~Courtoy, P.~Nadolsky, F.~Olness, and M.~Ponce-Chavez, ``{An
  analysis of parton distributions in a pion with B\'ezier parametrizations},''
  \href{http://arxiv.org/abs/2311.08447}{{\ttfamily arXiv:2311.08447
  [hep-ph]}}.

\bibitem{NNPDF:2021njg}
{\bfseries NNPDF} Collaboration, R.~D. Ball {\em et~al.}, ``{The path to proton
  structure at 1\% accuracy},''
  \href{http://dx.doi.org/10.1140/epjc/s10052-022-10328-7}{{\em Eur. Phys. J.
  C} {\bfseries 82} no.~5, (2022) 428},
  \href{http://arxiv.org/abs/2109.02653}{{\ttfamily arXiv:2109.02653
  [hep-ph]}}.

\bibitem{Courtoy:2022ocu}
A.~Courtoy, J.~Huston, P.~Nadolsky, K.~Xie, M.~Yan, and C.~P. Yuan, ``{Parton
  distributions need representative sampling},''
  \href{http://dx.doi.org/10.1103/PhysRevD.107.034008}{{\em Phys. Rev. D}
  {\bfseries 107} no.~3, (2023) 034008},
  \href{http://arxiv.org/abs/2205.10444}{{\ttfamily arXiv:2205.10444
  [hep-ph]}}.

\bibitem{Farina:2018fyg}
M.~Farina, Y.~Nakai, and D.~Shih, ``{Searching for New Physics with Deep
  Autoencoders},'' \href{http://dx.doi.org/10.1103/PhysRevD.101.075021}{{\em
  Phys. Rev. D} {\bfseries 101} no.~7, (2020) 075021},
  \href{http://arxiv.org/abs/1808.08992}{{\ttfamily arXiv:1808.08992
  [hep-ph]}}.

\bibitem{Cheng:2020dal}
T.~Cheng, J.-F. Arguin, J.~Leissner-Martin, J.~Pilette, and T.~Golling,
  ``{Variational autoencoders for anomalous jet tagging},''
  \href{http://dx.doi.org/10.1103/PhysRevD.107.016002}{{\em Phys. Rev. D}
  {\bfseries 107} no.~1, (2023) 016002},
  \href{http://arxiv.org/abs/2007.01850}{{\ttfamily arXiv:2007.01850
  [hep-ph]}}.

\bibitem{Finke:2021sdf}
T.~Finke, M.~Kr\"amer, A.~Morandini, A.~M\"uck, and I.~Oleksiyuk,
  ``{Autoencoders for unsupervised anomaly detection in high energy physics},''
  \href{http://dx.doi.org/10.1007/JHEP06(2021)161}{{\em JHEP} {\bfseries 06}
  (2021) 161}, \href{http://arxiv.org/abs/2104.09051}{{\ttfamily
  arXiv:2104.09051 [hep-ph]}}.

\bibitem{Ngairangbam:2021yma}
V.~S. Ngairangbam, M.~Spannowsky, and M.~Takeuchi, ``{Anomaly detection in
  high-energy physics using a quantum autoencoder},''
  \href{http://dx.doi.org/10.1103/PhysRevD.105.095004}{{\em Phys. Rev. D}
  {\bfseries 105} no.~9, (2022) 095004},
  \href{http://arxiv.org/abs/2112.04958}{{\ttfamily arXiv:2112.04958
  [hep-ph]}}.

\bibitem{Cerri:2018anq}
O.~Cerri, T.~Q. Nguyen, M.~Pierini, M.~Spiropulu, and J.-R. Vlimant,
  ``{Variational Autoencoders for New Physics Mining at the Large Hadron
  Collider},'' \href{http://dx.doi.org/10.1007/JHEP05(2019)036}{{\em JHEP}
  {\bfseries 05} (2019) 036}, \href{http://arxiv.org/abs/1811.10276}{{\ttfamily
  arXiv:1811.10276 [hep-ex]}}.

\bibitem{Govorkova:2021utb}
E.~Govorkova {\em et~al.}, ``{Autoencoders on field-programmable gate arrays
  for real-time, unsupervised new physics detection at 40 MHz at the Large
  Hadron Collider},'' \href{http://dx.doi.org/10.1038/s42256-022-00441-3}{{\em
  Nature Mach. Intell.} {\bfseries 4} (2022) 154--161},
  \href{http://arxiv.org/abs/2108.03986}{{\ttfamily arXiv:2108.03986
  [physics.ins-det]}}.

\bibitem{Hashimoto:2019bih}
K.~Hashimoto, ``{AdS/CFT correspondence as a deep Boltzmann machine},''
  \href{http://dx.doi.org/10.1103/PhysRevD.99.106017}{{\em Phys. Rev. D}
  {\bfseries 99} no.~10, (2019) 106017},
  \href{http://arxiv.org/abs/1903.04951}{{\ttfamily arXiv:1903.04951
  [hep-th]}}.

\bibitem{Hao:2022zns}
Z.~Hao, R.~Kansal, J.~Duarte, and N.~Chernyavskaya, ``{Lorentz group
  equivariant autoencoders},''
  \href{http://dx.doi.org/10.1140/epjc/s10052-023-11633-5}{{\em Eur. Phys. J.
  C} {\bfseries 83} no.~6, (2023) 485},
  \href{http://arxiv.org/abs/2212.07347}{{\ttfamily arXiv:2212.07347
  [hep-ex]}}.

\bibitem{Almaeen:2022ifg}
M.~Almaeen, Y.~Alanazi, N.~Sato, W.~Melnitchouk, and Y.~Li,
  \href{http://dx.doi.org/10.1109/ICMLA55696.2022.00187}{``{Point Cloud-based
  Variational Autoencoder Inverse Mappers (PC-VAIM) - An Application on Quantum
  Chromodynamics Global Analysis},''}
\newblock 12, 2022.

\bibitem{9534012}
M.~Almaeen, Y.~Alanazi, N.~Sato, W.~Melnitchouk, M.~P. Kuchera, and Y.~Li,
  \href{http://dx.doi.org/10.1109/IJCNN52387.2021.9534012}{``Variational
  autoencoder inverse mapper: An end-to-end deep learning framework for inverse
  problems,''} in {\em 2021 International Joint Conference on Neural Networks
  (IJCNN)}, pp.~1--8.
\newblock 2021.

\bibitem{Bringewatt:2020ixn}
J.~Bringewatt, N.~Sato, W.~Melnitchouk, J.-W. Qiu, F.~Steffens, and
  M.~Constantinou, ``{Confronting lattice parton distributions with global QCD
  analysis},'' \href{http://dx.doi.org/10.1103/PhysRevD.103.016003}{{\em Phys.
  Rev. D} {\bfseries 103} no.~1, (2021) 016003},
  \href{http://arxiv.org/abs/2010.00548}{{\ttfamily arXiv:2010.00548
  [hep-ph]}}.

\bibitem{Hou:2022onq}
T.-J. Hou, H.-W. Lin, M.~Yan, and C.~P. Yuan, ``{Impact of lattice strangeness
  asymmetry data in the CTEQ-TEA global analysis},''
  \href{http://dx.doi.org/10.1103/PhysRevD.107.076018}{{\em Phys. Rev. D}
  {\bfseries 107} no.~7, (2023) 076018},
  \href{http://arxiv.org/abs/2211.11064}{{\ttfamily arXiv:2211.11064
  [hep-ph]}}.

\bibitem{Detmold:2001dv}
W.~Detmold, W.~Melnitchouk, and A.~W. Thomas, ``{Parton distributions from
  lattice QCD},'' \href{http://dx.doi.org/10.1007/s1010501c0013}{{\em Eur.
  Phys. J. direct} {\bfseries 3} no.~1, (2001) 13},
  \href{http://arxiv.org/abs/hep-lat/0108002}{{\ttfamily
  arXiv:hep-lat/0108002}}.

\bibitem{Ji:2014gla}
X.~Ji, ``{Parton Physics from Large-Momentum Effective Field Theory},''
  \href{http://dx.doi.org/10.1007/s11433-014-5492-3}{{\em Sci. China Phys.
  Mech. Astron.} {\bfseries 57} (2014) 1407--1412},
  \href{http://arxiv.org/abs/1404.6680}{{\ttfamily arXiv:1404.6680 [hep-ph]}}.

\bibitem{Radyushkin:2017cyf}
A.~V. Radyushkin, ``{Quasi-parton distribution functions, momentum
  distributions, and pseudo-parton distribution functions},''
  \href{http://dx.doi.org/10.1103/PhysRevD.96.034025}{{\em Phys. Rev. D}
  {\bfseries 96} no.~3, (2017) 034025},
  \href{http://arxiv.org/abs/1705.01488}{{\ttfamily arXiv:1705.01488
  [hep-ph]}}.

\bibitem{Lin:2017snn}
H.-W. Lin {\em et~al.}, ``{Parton distributions and lattice QCD calculations: a
  community white paper},''
  \href{http://dx.doi.org/10.1016/j.ppnp.2018.01.007}{{\em Prog. Part. Nucl.
  Phys.} {\bfseries 100} (2018) 107--160},
  \href{http://arxiv.org/abs/1711.07916}{{\ttfamily arXiv:1711.07916
  [hep-ph]}}.

\bibitem{Constantinou:2020hdm}
M.~Constantinou {\em et~al.}, ``{Parton distributions and lattice-QCD
  calculations: Toward 3D structure},''
  \href{http://dx.doi.org/10.1016/j.ppnp.2021.103908}{{\em Prog. Part. Nucl.
  Phys.} {\bfseries 121} (2021) 103908},
  \href{http://arxiv.org/abs/2006.08636}{{\ttfamily arXiv:2006.08636
  [hep-ph]}}.

\bibitem{Smidt:2020tuy}
T.~E. Smidt, M.~Geiger, and B.~K. Miller, ``{Finding symmetry breaking order
  parameters with Euclidean neural networks},''
  \href{http://dx.doi.org/10.1103/PhysRevResearch.3.L012002}{{\em Phys. Rev.
  Res.} {\bfseries 3} no.~1, (2021) L012002},
  \href{http://arxiv.org/abs/2007.02005}{{\ttfamily arXiv:2007.02005 [cs.LG]}}.

\bibitem{Desai:2021wbb}
K.~Desai, B.~Nachman, and J.~Thaler, ``{Symmetry discovery with deep
  learning},'' \href{http://dx.doi.org/10.1103/PhysRevD.105.096031}{{\em Phys.
  Rev. D} {\bfseries 105} no.~9, (2022) 096031},
  \href{http://arxiv.org/abs/2112.05722}{{\ttfamily arXiv:2112.05722
  [hep-ph]}}.

\bibitem{Liu:2022plj}
D.~Liu, C.~Sun, and J.~Gao, ``{Machine learning of log-likelihood functions in
  global analysis of parton distributions},''
  \href{http://dx.doi.org/10.1007/JHEP08(2022)088}{{\em JHEP} {\bfseries 08}
  (2022) 088}, \href{http://arxiv.org/abs/2201.06586}{{\ttfamily
  arXiv:2201.06586 [hep-ph]}}.

\bibitem{Collins:2011zzd}
J.~Collins, \href{http://dx.doi.org/10.1017/9781009401845}{{\em {Foundations of
  perturbative QCD}}}, vol.~32.
\newblock Cambridge University Press, 11, 2013.

\bibitem{PDF4LHCWorkingGroup:2022cjn}
{\bfseries PDF4LHC Working Group} Collaboration, R.~D. Ball {\em et~al.},
  ``{The PDF4LHC21 combination of global PDF fits for the LHC Run III},''
  \href{http://dx.doi.org/10.1088/1361-6471/ac7216}{{\em J. Phys. G} {\bfseries
  49} no.~8, (2022) 080501}, \href{http://arxiv.org/abs/2203.05506}{{\ttfamily
  arXiv:2203.05506 [hep-ph]}}.

\bibitem{Butterworth:2015oua}
J.~Butterworth {\em et~al.}, ``{PDF4LHC recommendations for LHC Run II},''
  \href{http://dx.doi.org/10.1088/0954-3899/43/2/023001}{{\em J. Phys. G}
  {\bfseries 43} (2016) 023001},
  \href{http://arxiv.org/abs/1510.03865}{{\ttfamily arXiv:1510.03865
  [hep-ph]}}.

\bibitem{Ball:2012wy}
R.~D. Ball {\em et~al.}, ``{Parton Distribution Benchmarking with LHC Data},''
  \href{http://dx.doi.org/10.1007/JHEP04(2013)125}{{\em JHEP} {\bfseries 04}
  (2013) 125}, \href{http://arxiv.org/abs/1211.5142}{{\ttfamily arXiv:1211.5142
  [hep-ph]}}.

\bibitem{Bhattacharya:2023ays}
S.~Bhattacharya, K.~Cichy, M.~Constantinou, X.~Gao, A.~Metz, J.~Miller,
  S.~Mukherjee, P.~Petreczky, F.~Steffens, and Y.~Zhao, ``{Moments of proton
  GPDs from the OPE of nonlocal quark bilinears up to NNLO},''
  \href{http://dx.doi.org/10.1103/PhysRevD.108.014507}{{\em Phys. Rev. D}
  {\bfseries 108} no.~1, (2023) 014507},
  \href{http://arxiv.org/abs/2305.11117}{{\ttfamily arXiv:2305.11117
  [hep-lat]}}.

\bibitem{Bhattacharya:2023jsc}
S.~Bhattacharya {\em et~al.}, ``{Generalized Parton Distributions from Lattice
  QCD with Asymmetric Momentum Transfer: Axial-vector case},''
  \href{http://arxiv.org/abs/2310.13114}{{\ttfamily arXiv:2310.13114
  [hep-lat]}}.

\bibitem{Bhattacharya:2022aob}
S.~Bhattacharya, K.~Cichy, M.~Constantinou, J.~Dodson, X.~Gao, A.~Metz,
  S.~Mukherjee, A.~Scapellato, F.~Steffens, and Y.~Zhao, ``{Generalized parton
  distributions from lattice QCD with asymmetric momentum transfer: Unpolarized
  quarks},'' \href{http://dx.doi.org/10.1103/PhysRevD.106.114512}{{\em Phys.
  Rev. D} {\bfseries 106} no.~11, (2022) 114512},
  \href{http://arxiv.org/abs/2209.05373}{{\ttfamily arXiv:2209.05373
  [hep-lat]}}.

\bibitem{Ji:1996ek}
X.-D. Ji, ``{Gauge-Invariant Decomposition of Nucleon Spin},''
  \href{http://dx.doi.org/10.1103/PhysRevLett.78.610}{{\em Phys. Rev. Lett.}
  {\bfseries 78} (1997) 610--613},
  \href{http://arxiv.org/abs/hep-ph/9603249}{{\ttfamily arXiv:hep-ph/9603249}}.

\bibitem{Muller:1994ses}
D.~M\"uller, D.~Robaschik, B.~Geyer, F.~M. Dittes, and J.~Ho\v{r}ej\v{s}i,
  ``{Wave functions, evolution equations and evolution kernels from light ray
  operators of QCD},'' \href{http://dx.doi.org/10.1002/prop.2190420202}{{\em
  Fortsch. Phys.} {\bfseries 42} (1994) 101--141},
  \href{http://arxiv.org/abs/hep-ph/9812448}{{\ttfamily arXiv:hep-ph/9812448}}.

\bibitem{Radyushkin:1997ki}
A.~V. Radyushkin, ``{Nonforward parton distributions},''
  \href{http://dx.doi.org/10.1103/PhysRevD.56.5524}{{\em Phys. Rev. D}
  {\bfseries 56} (1997) 5524--5557},
  \href{http://arxiv.org/abs/hep-ph/9704207}{{\ttfamily arXiv:hep-ph/9704207}}.

\bibitem{Ji:1996nm}
X.-D. Ji, ``{Deeply virtual Compton scattering},''
  \href{http://dx.doi.org/10.1103/PhysRevD.55.7114}{{\em Phys. Rev. D}
  {\bfseries 55} (1997) 7114--7125},
  \href{http://arxiv.org/abs/hep-ph/9609381}{{\ttfamily arXiv:hep-ph/9609381}}.

\bibitem{Ji:1997gm}
X.-D. Ji, W.~Melnitchouk, and X.~Song, ``{A Study of off forward parton
  distributions},'' \href{http://dx.doi.org/10.1103/PhysRevD.56.5511}{{\em
  Phys. Rev. D} {\bfseries 56} (1997) 5511--5523},
  \href{http://arxiv.org/abs/hep-ph/9702379}{{\ttfamily arXiv:hep-ph/9702379}}.

\bibitem{Diehl:2003ny}
M.~Diehl, ``{Generalized parton distributions},''
  \href{http://dx.doi.org/10.1016/j.physrep.2003.08.002}{{\em Phys. Rept.}
  {\bfseries 388} (2003) 41--277},
  \href{http://arxiv.org/abs/hep-ph/0307382}{{\ttfamily arXiv:hep-ph/0307382}}.

\bibitem{37f2b6bee745402aa4e4d124d33be0e0}
Y.~Lecun, {\em PhD thesis: Modeles connexionnistes de l'apprentissage
  (connectionist learning models)}.
\newblock Universite P. et M. Curie (Paris 6), June, 1987.

\bibitem{Bourlard1988AutoassociationBM}
H.~Bourlard and Y.~Kamp, ``Auto-association by multilayer perceptrons and
  singular value decomposition,'' {\em Biological Cybernetics} {\bfseries 59}
  (1988) 291--294. \url{https://api.semanticscholar.org/CorpusID:206775335}.

\bibitem{Goodfellow-et-al-2016}
I.~Goodfellow, Y.~Bengio, and A.~Courville, {\em Deep Learning}.
\newblock MIT Press, 2016.
\newblock \url{http://www.deeplearningbook.org}.

\bibitem{agarap2019deep}
A.~F. Agarap, ``Deep learning using rectified linear units (relu),'' 2019.

\bibitem{clevert2016fast}
D.-A. Clevert, T.~Unterthiner, and S.~Hochreiter, ``Fast and accurate deep
  network learning by exponential linear units (elus),'' 2016.

\bibitem{10.5555/2832747.2832823}
F.~Zhuang, X.~Cheng, P.~Luo, S.~J. Pan, and Q.~He, ``Supervised representation
  learning: Transfer learning with deep autoencoders,'' in {\em Proceedings of
  the 24th International Conference on Artificial Intelligence}, IJCAI'15,
  p.~4119–4125.
\newblock AAAI Press, 2015.

\bibitem{doi:10.1126/science.1127647}
G.~E. Hinton and R.~R. Salakhutdinov, ``Reducing the dimensionality of data
  with neural networks,'' \href{http://dx.doi.org/10.1126/science.1127647}{{\em
  Science} {\bfseries 313} no.~5786, (2006) 504--507},
  \href{http://arxiv.org/abs/https://www.science.org/doi/pdf/10.1126/science.1127647}{{\ttfamily
  https://www.science.org/doi/pdf/10.1126/science.1127647}}.
  \url{https://www.science.org/doi/abs/10.1126/science.1127647}.

\bibitem{WANG2016232}
Y.~Wang, H.~Yao, and S.~Zhao, ``Auto-encoder based dimensionality reduction,''
  \href{http://dx.doi.org/https://doi.org/10.1016/j.neucom.2015.08.104}{{\em
  Neurocomputing} {\bfseries 184} (2016) 232--242}.
  \url{https://www.sciencedirect.com/science/article/pii/S0925231215017671}.
  RoLoD: Robust Local Descriptors for Computer Vision 2014.

\bibitem{10.1145/1390156.1390294}
P.~Vincent, H.~Larochelle, Y.~Bengio, and P.-A. Manzagol,
  \href{http://dx.doi.org/10.1145/1390156.1390294}{``Extracting and composing
  robust features with denoising autoencoders,''} ICML '08, p.~1096–1103.
\newblock Association for Computing Machinery, New York, NY, USA, 2008.
\newblock \url{https://doi.org/10.1145/1390156.1390294}.

\bibitem{8363930}
Z.~Chen, C.~K. Yeo, B.~S. Lee, and C.~T. Lau,
  \href{http://dx.doi.org/10.1109/WTS.2018.8363930}{``Autoencoder-based network
  anomaly detection,''} in {\em 2018 Wireless Telecommunications Symposium
  (WTS)}, pp.~1--5.
\newblock 2018.

\bibitem{kingma2022autoencoding}
D.~P. Kingma and M.~Welling, ``Auto-encoding variational bayes,'' 2022.

\bibitem{DBLP:journals/corr/abs-1906-02691}
D.~P. Kingma and M.~Welling, ``An introduction to variational autoencoders,''
  {\em CoRR} {\bfseries abs/1906.02691} (2019) ,
  \href{http://arxiv.org/abs/1906.02691}{{\ttfamily 1906.02691}}.
  \url{http://arxiv.org/abs/1906.02691}.

\bibitem{DBLP:journals/corr/HeZRS15}
K.~He, X.~Zhang, S.~Ren, and J.~Sun, ``Deep residual learning for image
  recognition,'' {\em CoRR} {\bfseries abs/1512.03385} (2015) ,
  \href{http://arxiv.org/abs/1512.03385}{{\ttfamily 1512.03385}}.
  \url{http://arxiv.org/abs/1512.03385}.

\bibitem{DBLP:journals/corr/abs-1205-2653}
C.~Cortes, M.~Mohri, and A.~Rostamizadeh, ``{L2} regularization for learning
  kernels,'' {\em CoRR} {\bfseries abs/1205.2653} (2012) ,
  \href{http://arxiv.org/abs/1205.2653}{{\ttfamily 1205.2653}}.
  \url{http://arxiv.org/abs/1205.2653}.

\bibitem{DBLP:journals/corr/abs-2011-06225}
M.~Abdar, F.~Pourpanah, S.~Hussain, D.~Rezazadegan, L.~Liu, M.~Ghavamzadeh,
  P.~W. Fieguth, X.~Cao, A.~Khosravi, U.~R. Acharya, V.~Makarenkov, and
  S.~Nahavandi, ``A review of uncertainty quantification in deep learning:
  Techniques, applications and challenges,'' {\em CoRR} {\bfseries
  abs/2011.06225} (2020) , \href{http://arxiv.org/abs/2011.06225}{{\ttfamily
  2011.06225}}. \url{https://arxiv.org/abs/2011.06225}.

\bibitem{MacKay1995ProbableNA}
D.~J.~C. MacKay, ``Probable networks and plausible predictions - a review of
  practical bayesian methods for supervised neural networks,'' {\em Network:
  Computation In Neural Systems} {\bfseries 6} (1995) 469--505.
  \url{https://api.semanticscholar.org/CorpusID:14332165}.

\bibitem{srivastava2014dropout}
N.~Srivastava, G.~Hinton, A.~Krizhevsky, I.~Sutskever, and R.~Salakhutdinov,
  ``Dropout: a simple way to prevent neural networks from overfitting,'' {\em
  The journal of machine learning research} {\bfseries 15} no.~1, (2014)
  1929--1958.

\bibitem{gal2016dropout}
Y.~Gal and Z.~Ghahramani, ``Dropout as a bayesian approximation: Representing
  model uncertainty in deep learning,'' 2016.

\bibitem{malinin2018predictive}
A.~Malinin and M.~Gales, ``Predictive uncertainty estimation via prior
  networks,'' 2018.

\bibitem{Almaeen:2022imx}
M.~Almaeen, J.~Grigsby, J.~Hoskins, B.~Kriesten, Y.~Li, H.-W. Lin, and
  S.~Liuti, ``{Benchmarks for a Global Extraction of Information from Deeply
  Virtual Exclusive Scattering},''
  \href{http://arxiv.org/abs/2207.10766}{{\ttfamily arXiv:2207.10766
  [hep-ph]}}.

\bibitem{Hunt-Smith:2022ugn}
N.~T. Hunt-Smith, A.~Accardi, W.~Melnitchouk, N.~Sato, A.~W. Thomas, and M.~J.
  White, ``{Determination of uncertainties in parton densities},''
  \href{http://dx.doi.org/10.1103/PhysRevD.106.036003}{{\em Phys. Rev. D}
  {\bfseries 106} no.~3, (2022) 036003},
  \href{http://arxiv.org/abs/2206.10782}{{\ttfamily arXiv:2206.10782
  [hep-ph]}}.

\bibitem{Guzzi:2022rca}
M.~Guzzi, T.~J. Hobbs, K.~Xie, J.~Huston, P.~Nadolsky, and C.~P. Yuan, ``{The
  persistent nonperturbative charm enigma},''
  \href{http://dx.doi.org/10.1016/j.physletb.2023.137975}{{\em Phys. Lett. B}
  {\bfseries 843} (2023) 137975},
  \href{http://arxiv.org/abs/2211.01387}{{\ttfamily arXiv:2211.01387
  [hep-ph]}}.

\bibitem{Hobbs:2013bia}
T.~J. Hobbs, J.~T. Londergan, and W.~Melnitchouk, ``{Phenomenology of
  nonperturbative charm in the nucleon},''
  \href{http://dx.doi.org/10.1103/PhysRevD.89.074008}{{\em Phys. Rev. D}
  {\bfseries 89} no.~7, (2014) 074008},
  \href{http://arxiv.org/abs/1311.1578}{{\ttfamily arXiv:1311.1578 [hep-ph]}}.

\bibitem{Hou:2019qau}
T.-J. Hou {\em et~al.}, ``{Progress in the CTEQ-TEA NNLO global QCD
  analysis},'' \href{http://arxiv.org/abs/1908.11394}{{\ttfamily
  arXiv:1908.11394 [hep-ph]}}.

\bibitem{Bailey:2020ooq}
S.~Bailey, T.~Cridge, L.~A. Harland-Lang, A.~D. Martin, and R.~S. Thorne,
  ``{Parton distributions from LHC, HERA, Tevatron and fixed target data:
  MSHT20 PDFs},'' \href{http://dx.doi.org/10.1140/epjc/s10052-021-09057-0}{{\em
  Eur. Phys. J. C} {\bfseries 81} no.~4, (2021) 341},
  \href{http://arxiv.org/abs/2012.04684}{{\ttfamily arXiv:2012.04684
  [hep-ph]}}.

\end{thebibliography}\endgroup

\end{document}